\documentclass[traditabstract]{aa} 
\usepackage{graphicx}
\usepackage{txfonts}

\usepackage{setspace}
\newcommand{\INT}{{\it INTEGRAL}\,}
\usepackage{natbib}

\begin{document}
   \title{The spectral catalogue of {\it INTEGRAL} gamma-ray bursts:}

   \subtitle{results of the joint IBIS/SPI spectral analysis}
   \author{\v Z. Bo\v snjak
          \inst{1,2,3}
          \and
          D. G\"otz\inst{1}
          \and
          L. Bouchet\inst{3,4}
          \and
          S. Schanne\inst{1}
          \and
          B. Cordier\inst{1}
          }

   \institute{AIM (UMR 7158 CEA/DSM-CNRS-Universit\'e Paris Diderot) Irfu/Service d'Astrophysique, Saclay, F-91191 Gif-sur-Yvette Cedex, France\\
              \email{zeljka.bosnjak@cea.fr}
             \and
 Department of Physics, University of Rijeka, 51000 Rijeka, Croatia
             \and
 Universit\'e de Toulouse, UPS-OMP, IRAP, Toulouse, France
             \and
 CNRS, IRAP, 9 Av. Colonel Roche, BP 44346, F-31028 Toulouse Cedex 4, France
             }

   \date{Received September ; accepted }

\abstract 
{We present the updated {\it INTEGRAL} catalogue of gamma-ray bursts (GRBs) observed between December 2002 and February 2012. The catalogue contains the spectral parameters for 59 GRBs localized by the {\it INTEGRAL} Burst Alert System (IBAS). We used the data from the two main instruments on board the {\it INTEGRAL} satellite: the spectrometer SPI (SPectrometer on {\it INTEGRAL}) nominally covering the energy range 18 keV -- 8 MeV, and the imager IBIS (the Imager on Board the {\it INTEGRAL} Satellite) operating in the range from 15 keV to 10 MeV.  For the spectral analysis we applied a new data extraction technique, developed in order to explore the energy regions of highest sensitivity for both instruments, SPI and IBIS. It allowed us to perform analysis of the GRB spectra over a broad energy range and to determine the bursts' spectral peak energies. The spectral analysis was performed on the whole sample of GRBs triggered by IBAS, including all the events observed in period December 2002 -- February 2012. The catalogue contains the trigger times, burst coordinates, positional errors, durations and peak fluxes for 28 unpublished GRBs observed between September 2008 and February 2012. The light curves in 20 -- 200 keV energy band of these events were derived using IBIS data. We compare the prompt emission properties of the {\it INTEGRAL} GRB sample with the BATSE and {\it Fermi} samples.}


   \maketitle
%

\section{Introduction}

Over the past two decades Gamma-Ray Bursts (GRBs) have been observed by several missions, providing a wealth of spectral and temporal data. The properties of the prompt gamma-ray emission have been studied over a broad energy range, from keV to GeV energies. Prompt GRB spectra are commonly described by two power laws smoothly connected around the spectral peak energy typically observed at a few hundred keV \citep{band93,preece00}.  The values of the spectral parameters, i.e. the slopes of the low- and high-energy power laws and the peak energy, are associated with the radiative mechanisms governing the emission and with the energy dissipation processes within the relativistic jet, and therefore impose important constraints on the theoretical models for prompt GRB emission. To date the most complete catalogues of spectral GRB properties comprise the events observed by BATSE (Burst And Transient Source Experiment) on board the {\it Compton Gamma Ray Observatory} in operation from 1991 to 2000 \citep{fishman89,gehrels94}, by the {\it Swift} satellite launched in 2004 \citep{gehrels04}, and by the {\it Fermi} satellite launched in 2008 \citep{meegan09,gehrels13}.

 {\it INTEGRAL} has contributed important discoveries to the GRB field, including the detection and observation of GRB 031203 associated with SN 2003lw \citep{malesani04}, the polarization measurements from GRB 041219 \citep{gotz09} and GRB 061122 \citep{gotz13}, and the discovery of the (inferred) population of low-luminosity GRBs \citep{foley08}. In this paper we present a catalogue of gamma-ray bursts detected by the {\it INTEGRAL} satellite. In the period between December 2002 and February 2012 {\it INTEGRAL} observed 83 GRBs (the low number of events with respect to other GRB missions, e.g. 2700 GRBs observed by BATSE in a nine year period, is mainly due to the small field of view of the IBIS instrument, $\sim$ 0.1 sr). We report the results of spectral analysis of 59 out of 83 GRBs. The spectral parameters were derived by combining the data from the two main instruments on board {\it INTEGRAL}, the spectrometer SPI nominally covering the energy range 18 keV - 8 MeV, and the imager IBIS with spectral sensitivity in the range 15 keV - 10 MeV. To date, the systematic spectral analysis of {\it INTEGRAL} GRBs has been performed in a limited energy range using only the data from the IBIS instrument \citep{vianello09,foley08,tierney10}. \citet{foley08} reported the results of the spectral analysis using SPI data for nine GRBs, but, with one exception, the analysis of the IBIS data and the SPI data has been performed independently. In addition to the spectral analysis performed over a broad energy range for the complete sample of {\it INTEGRAL} GRBs, we have derived the IBIS light curves and durations for the previously unpublished 28 events observed between September 2008 and February 2012.

The paper is organised as follows. In Section 2 we discuss the catalogues of GRBs detected by BATSE, {\it Fermi} and {\it Swift}, and the possible biases in the results due to the instrumental differences. We compare the instrumental properties of different missions with those of the {\it INTEGRAL} instruments. The timing analysis and the spectral extraction technique we developed are presented in Section 3. The basic properties of {\it INTEGRAL} GRB sample are discussed in Section 4; we compare the global properties of our sample with the large GRB samples obtained by {\it CGRO} BATSE, {\it Fermi} Gamma-Ray Burst Monitor (GBM) and {\it Swift} Burst Alert Telescope (BAT) instruments. We report the results of the spectral analysis in Section 5, and make a statistical comparison of our results with respect to BATSE and {\it Fermi}/GBM samples. The summary of our results is presented in Section 6.

\section{GRB samples}

 A systematic spectral analysis of a sub-sample of 350 bright GRBs selected from the complete set of 2704 BATSE GRBs, on the energy range $\sim$ 30 keV - 2 MeV observed by BATSE was performed by \citet{kaneko06} (see also \citealt{preece00}). Five percent of the bursts in this sample are short GRBs (with durations less than 2 seconds). \citet{kaneko06} found that the most common value for the low-energy slope of the photon spectra is $\alpha \sim$ --1, and therefore the distribution of the low-energy indices is not consistent with the value predicted by the standard synchrotron emission model in fast cooling regime, --3/2 \citep{sari98}. The distribution of the peak energies of the time-integrated spectra of long BATSE GRBs has a maximum at $\sim$ 250 keV, and a very narrow width $\lesssim$ 100 keV.
\citet{ghirlanda04} found for a sample of short GRBs observed by BATSE that their time-integrated spectra are harder than those of long GRBs spectra, mainly due to a harder low-energy spectral component ($\sim$ --0.6). \citet{goldstein12} (see also \citealt{bissaldi11}, \citealt{nava11}) reported the results of the spectral analysis of 487 GRBs detected by {\it Fermi}/GBM operating in the energy band $\sim$ 8 keV -- 40 MeV, during its first two years of operation. They found that the distribution of spectral peak energies has a maximum at $\sim$ 200 keV for the time integrated spectra, and also reported several GRBs with time-integrated peak energies greater than 1 MeV. The properties of the sample of 476 GRBs observed by {\it Swift}/BAT on 15--150 keV energy range were reported by \citet{sakamoto11}. They distinguish the classes of long duration GRBs (89\%), short duration GRBs (8\%), and short-duration GRBs with extended emission (2\%). Their GRB sample was found to be significantly softer than the BATSE bright GRBs, having time-integrated peak energies around $\sim$ 80 keV.

To test the emission models using the observed spectral parameters or to deduce some global properties of a GRB sample, it is necessary to take into account the possible biases in the results of the spectral analysis:
\begin{enumerate} 

\item When we consider the low energy portion of the spectrum, the data may not approach low energy asymptotic power law within the energy band of the instrument \citep{preece98,kaneko06}: e.g. if the spectral peak energy is close to the lower edge of the instrumental energy band, lower values of $\alpha$ are determined. \citet{preece98} introduced as a better measure of the low energy spectral index the effective value of $\alpha$, defined as the slope of the power law tangent to the GRB spectrum at some chosen energy (25 keV for BATSE data). 

\item There may be biases in the results when the analysis is performed only on a sample of the brightest events, since there is a tendency for bright GRBs (having higher photon fluxes) to have higher spectral peak energies than faint GRBs \citep{borgonovo01,mallozzi95}. For example, \citet{kaneko06} burst selection criteria required a peak photon flux on energy range 50--300 keV greater than 10 photons s$^{-1}$ cm$^{-2}$ or a total energy fluence in 20--2000 keV energy range larger than 2.0 $\times$ 10$^{-5}$ erg cm$^{-2}$. \citet{nava08} have extended the spectral analysis of BATSE GRBs to the fainter bursts (down to fluences $\sim$ 10$^{-6}$ erg cm$^{-2}$) and found a lower value for the average spectral peak energy, $\sim$ 160 keV, with respect to the \citet{kaneko06} results.

\item The instrumental selection effects (e.g. the integration time scale for the burst trigger) may also affect the properties of the GRB samples obtained by different gamma-ray experiments. For example, \citet{sakamoto11} (see \citealt{qin13} for {\it Fermi}/GBM results) found that the distribution of long GRB durations from {\it Swift}/BAT sample is shifted towards longer times ($\sim$ 70 s) with respect to BATSE ($\sim$ 25 s, \citealt{kouveliotou93}), coherently with the longer BAT triggering time scales. The lack of short GRBs in imaging instruments (such as {\it Swift}/BAT) with respect to non-imaging instruments (like BATSE and GBM) on the other hand, is attributed to the requirement for a minimum number of photons needed to build an image with a coded-mask instrument. 

\end{enumerate}

\subsection{{\it INTEGRAL} instruments}

\INT \citep{winkler03} is an ESA mission launched on October 17, 2002 dedicated to high resolution imaging and spectroscopy in the hard X-/soft $\gamma$-ray domain.
It carries two main coded-mask instruments, SPI \citep{vedrenne03}, and IBIS \citep{ubertini03}. 

SPI is made of 19 Ge detectors\footnote{During the mission lifetime four SPI detectors have failed, and this has been accounted for in our spectral analysis.}, working in the 20 keV--8 MeV energy range, and is optimized for high resolution spectroscopy ($\sim$2 keV @ 1 MeV), in spite of a relatively poor spatial resolving power of $\sim$2$^{\circ}$. IBIS is made of two pixellated detection planes: the upper plane, ISGRI -- {\it INTEGRAL} Soft Gamma-Ray Imager \citep{lebrun03}, is made of 128$\times$128 CdTe detectors and operates in the 15 keV--1 MeV energy range. ISGRI has an unprecedented point spread function (PSF) in the soft $\gamma$-ray domain of 12 arc min FWHM. The lower detection plane, PICsIT -- PIxellated CsI Telescope \citep{dicocco03}, is made of 64$\times$64 pixels of CsI, and is sensitive between 150 keV and 10 MeV. For our analysis we used only ISGRI data from IBIS. Due to satellite telemetry limitations, PICsIT data are temporally binned over the duration of \INT pointing (lasting typically 30--45 minutes) and hence they are not suited for studies of short transients like GRBs.

Despite being a non GRB-oriented mission, \INT can be used as a GRB experiment: the GRBs presented in this paper have all been detected in (near-)real time by the \INT Burst Alert System (IBAS; \citealt{mereghetti03}). IBAS is running on ground at the INTEGRAL Science Data Centre (ISDC; \citealt{courvoisier03}) thanks to the continuous downlink of the \INT  telemetry. As soon as the IBIS/ISGRI data are received at ISDC, they are analysed in real-time by several triggering processes running in parallel. The triggering algorithms are of two kinds: one is continuously comparing the current sky image with a reference image to look for new sources, and the second one examines the global count rate of ISGRI. In the latter case, once a significant excess is found, imaging is used to check if it corresponds to a new point source or if it has a different origin (e.g. cosmic rays or solar flares). The nominal triggering energy band for IBAS is 15--200 keV, and different time scales are explored, from 2 ms up to 100 s. 

\begin{figure}
\centering
\includegraphics[width=0.37\textwidth,angle=+90]{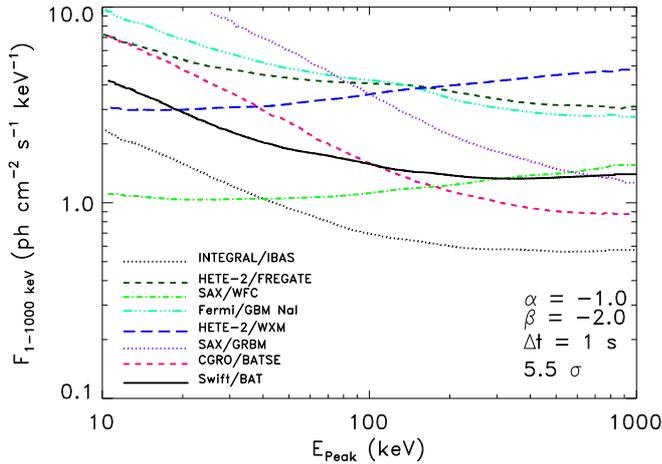}
\caption{Sensitivity of past and present GRB experiments as a function of the GRB peak energy. For {\it Fermi}/GBM NaI detectors, {\it BeppoSAX}/WFC, {\it BeppoSAX}/GRBM, {\it HETE-2}/Fregate, {\it HETE-2}/WXM, and {\it CGRO}/BATSE LAD detectors the data are taken from \citet{band03}; for {\it Swift}/BAT the data are taken from \citet{band06}; for {\it INTEGRAL}/IBAS the data are taken from the IBAS web site$^3$.}
         \label{fig:sensi}
\end{figure}

As a comparison, in Fig. \ref{fig:sensi} we show the sensitivity of some past and current GRB triggering experiments\footnote{For updated sensitivity of {\it Fermi}/GBM, see \citet{bissaldi09} and \citet{meegan09}.}, as a function of the GRB peak energy. The sensitivity is presented as the threshold peak photon flux (in 1--1000 keV energy band) detected at a 5.5 $\sigma$ signal-to-noise ratio, during accumulation time $\Delta$t = 1 s (cf. \citealt{band03}). It can be seen that the IBAS system is expected to be the most sensitive experiment provided that the peak energy is larger than $\sim$ 50 keV. This has allowed us to investigate the GRB spectral properties for faint (down to fluences of a few 10$^{-8}$ erg cm$^{-2}$) GRBs, see Fig. \ref{fig:fluence}.

\section{Data analysis}

\label{sec:analysis}

Due to their short durations, GRBs usually do not provide a large number of counts, especially above $\sim$ 200 keV where the IBIS/ISGRI effective area starts to decline rapidly. In order to provide a broader energy coverage and a better sensitivity for the \INT GRB spectra, we combined the data from the IBIS/ISGRI and the SPI instruments. 

The previously published catalogue of \INT GRBs \citep{vianello09} comprises the analysis of the IBIS/ISGRI data only, providing the GRB spectra on 18--300 keV band. Most of the spectra were fitted with a single power law model over this limited energy range. The spectral energy peak was determined for only 9 out of 56 bursts in their sample. \citet{foley08} used SPI and IBIS/ISGRI data to analyse 9 out of 45 GRBs. They performed the spectral analysis using the data from each of the two instruments independently; in four cases the spectrum was fitted by a different model for IBIS and for SPI data. 

\begin{figure*}[ht!]
\begin{center}
\centering
\includegraphics[width=0.5\textwidth,angle=-90]{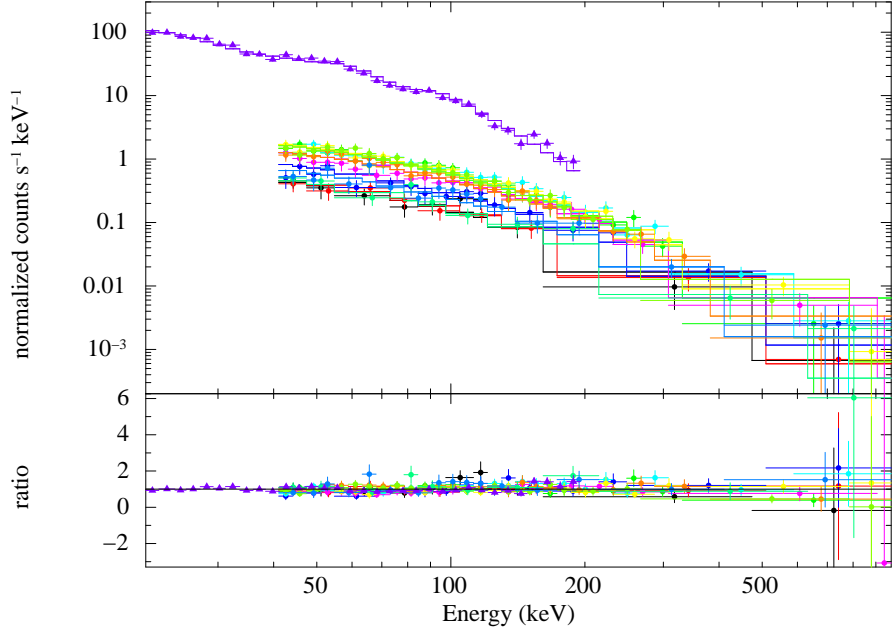}
\caption{The spectral model and the data for GRB 061122. The violet triangles are the 20-200 keV ISGRI data points used for the fit. The coloured points covering the energy range 40 - 1000 keV are the data from the 11 SPI detectors with the highest signal-to-noise ratio selected for the spectral analysis (cf. \citealt{mcglynn09}). We fitted the cut-off power law model in this case, with $\alpha$ = -1.3 and E$_0$ = 263 keV (see Section \ref{sec:spec} on spectral analysis).} 
         \label{fig:GRB061122}
\end{center}
\end{figure*}

We combined the data from both instruments, SPI and IBIS/ISGRI: in this way the low energy portion ($\lesssim$ 200 keV) of the GRB spectrum is explored by ISGRI where its sensitivity is highest, while the high energy portion of the spectrum ($\gtrsim$ 200 keV) is better investigated by the SPI data. Joint spectral analysis, using the data from both instruments (see Fig. \ref{fig:GRB061122}), allows us to analyze the spectra consistently and to exploit the maximum potential of each instrument. The SPI data can provide better spectral information at energies where IBIS/ISGRI effective area becomes low, and therefore are suitable to determine the GRB spectral peak energy (typically at $\sim$ a few 100 keV).

The spectra were analysed using the C-statistic \citep{cash79}, which is commonly used for experiments with a low number of counts. In order to fit the spectra of both instruments simultaneously, we used the XPSEC 12.7.1 fitting package \citep{arnaud10}. For the C-statistic to be applied, we needed to provide on-burst spectra and background spectra separately for every GRB. This cannot be obtained by the \INT standard Off-line Scientifc Analysis software (OSA), and therefore we developed additional tools to extract the spectra in the required format.

In order to maximize the sensitivity of both instruments, SPI and ISGRI spectra were extracted in the range (40 keV - 1 MeV) and (20 - 200 keV) respectively, because outside this energy range, the effective areas of the corresponding instruments decrease very rapidly (see \citealt{ubertini03,vedrenne03}).  First, we computed the ISGRI light curves for each GRB (Figs. A.1 - A.5). To enhance the signal-to-noise ratio, we selected only the events that were recorded by the pixels having more than 60\% of their surface illuminated by the GRB. 

Based on these light curves we selected on-burst intervals for spectral analysis (see dashed lines in Figs. A.1 - A.5). The off-burst intervals for the spectral analysis were determined by selecting the times before and after the GRB, excluding intervals of $\gtrsim$ 10 s close to the event in order to ensure that the off-burst intervals were not contaminated by the GRB counts. For the SPI instrument, a spectrum for each of the 19 (where applicable) Ge detectors was computed. The net individual GRB spectra (i.e. on-burst -- off-burst spectra) have the advantage (with respect to the \textit{global} spectra produced by OSA software) of being more accurate since the background spectra were computed for each GRB and each detector, taking into account the local spectral and temporal background evolution. The OSA software, on the contrary, computes the SPI background from a model template and the net spectrum is derived from the sky deconvolution process, which introduces more uncertainties than a simple subtraction of the number of background counts per detector. For each SPI detector an individual response function was calculated, taking into account the GRB direction (either as determined by IBIS/ISGRI or by more precise X-ray or optical follow-up observations). The response function takes into account the exposed fraction of each detector given the GRB direction. This means that for detectors that are completely shadowed by the SPI mask the corresponding net spectrum is consistent with zero (see e.g. \citealt{mcglynn09}), and it is automatically neglected in the analysis since the effective area is also consistent with zero.
For the IBIS/ISGRI spectra, due to the large number of detectors, we decided not to compute individual pixel spectra. We selected only the pixels that were fully illuminated by the GRB, in order to compute the off-burst and on-burst spectra. A corresponding ancillary response function (ARF) was computed, taking into account the reduced ($\sim$ 30\%) area of the detector plane we used. For each GRB we computed and fitted the time-integrated spectrum, using all the available SPI spectra and one ISGRI spectrum. In order to account for SPI/IBIS inter-calibration and especially for IBIS count losses due to telemetry limitations (see e.g. Fig. A.1 last panel), we allowed a constant normalization factor between ISGRI and SPI. On the other hand we assumed that the differences among the individual SPI detectors are all accounted for by the ad-hoc generated response matrices. An example of a simultaneous fit is shown in Fig. \ref{fig:GRB061122}.

\begin{figure}
\centering
\includegraphics[width=0.5\textwidth]{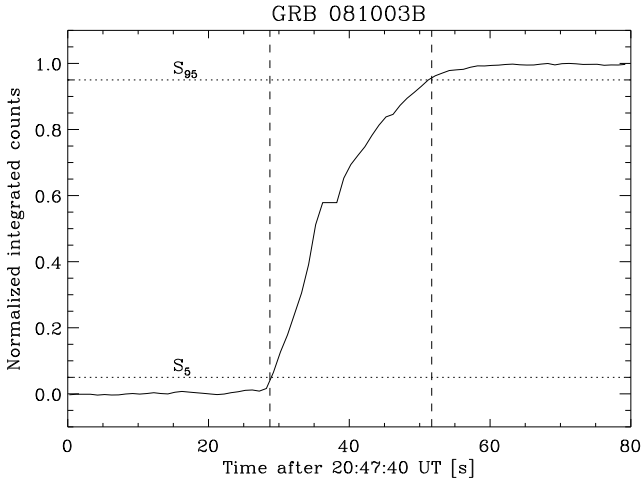}
\caption{An example of T$_{90}$ calculation using the time integrated IBIS/ISGRI light curve of GRB 081003B. The dashed vertical lines represent the times when the GRB integrated counts exceed the 5\% (S$_{5}$) and 95\% (S$_{95}$) of the maximal integrated flux value, respectively.}
         \label{fig:t90}
\end{figure}

We determined the T$_{90}$ duration for sample of GRBs observed after September 2008 (Table \ref{tab:t90}). The T$_{90}$ duration of prompt emission measures the duration of the time interval during which 90\% of the observed counts are accumulated \citep{kouveliotou93}. The start and the end of this interval are defined by the time at which 5\% and 95\% of the counts are accumulated (see Fig. \ref{fig:t90}), respectively. 

The GRB durations were determined using the IBIS/ISGRI light curves (see Figs. A.1 - A.5) obtained for 20--200 keV energy band. The background rate was determined by fitting a linear or constant function to the data, in the time intervals before and after burst. The time intervals for background fitting lasted typically for $\gtrsim$100 s, and they were separated by $\sim$10 s from the burst.  We show the background-subtracted ligthcurves for GRBs detected between September 2008 and February 2012 in Figs. A.1 - A.5. The errors on T$_{90}$ were calculated using the method developed by \citet{koshut96}. They define the total net (i.e. background subtracted) source counts observed for a single event as

\begin{equation}
S_{tot}=\int_{-\infty}^{+\infty} \frac{dS}{dt}dt
\end{equation}
where $dS/dt$ is the source count rate history. The time $\tau_{f}$ during which a given fraction $f$ of the total counts is accumulated is defined as the time at which

\begin{equation}
\frac{\int_{-\infty}^{\tau_{f}} \frac{dS}{dt} dt} {S_{tot}} = f.
\end{equation}
Hence T$_{90}$ is defined as $\tau_{95}-\tau_{5}$, see Fig. \ref{fig:t90}. $S_{f}=\int_{-\infty}^{\tau_{f}} \frac{dS}{dt} dt$ represents the value of the integrated counts $S(t)$ when $f$ of the total counts have been detected. The uncertainties on $S_{f}$ consist of two contributions: $(dS_{f})_{cnt}$, due to the uncertainty in the integrated counts $S(t)$ at any time $t$ (see Eq. (12) in \citealt{koshut96}), and $(dS_{f})_{fluc}$, due to the statistical fluctuations (see Eq. (14) in \citealt{koshut96}) with respect to the smooth background model:

\begin{equation}
 (dS_{f})_{tot}=\sqrt{(dS_{f})_{cnt}^{2}+(dS_{f})_{fluc}^{2}}. 
\end{equation}
The times $\tau_{f-}$ and $\tau_{f+}$ are the times at which $S_{f}-(dS_{f})_{tot}$ and 
$S_{f}+(dS_{f})_{tot}$ counts have been reached respectively. In this case, one can define

\begin{equation}
\Delta\tau_{f}=\tau_{f+}-\tau_{f-}
\end{equation}
and the statistical uncertainty in T$_{90}$ is given by

\begin{equation}
\delta T_{90}=\sqrt{(\Delta\tau_{5})^{2}+(\Delta\tau_{95})^{2}}.
\end{equation}

In Table \ref{tab:t90}, we report also the peak fluxes of the GRBs. They were calculated over 1 s for long GRBs, using the standard OSA v10.0 software and IBIS/ISGRI data only. We did not use our new spectral extraction method since due to the low statistics over such a short time interval the spectral fitting does not require more sophisticated models than a simple power law to estimate the peak flux. In this case, the large IBIS/ISGRI effective area is adapted to provide a fair measure of the peak flux. We also recalculated, using the latest available calibration files, the positions and the associated 90\% c.l. errors. They were computed using ISGRI only, due to its higher positional accuracy with respect to SPI.

\section{{\it INTEGRAL} GRB sample: global properties}

\begin{figure}
\includegraphics[width=0.5\textwidth,height=9cm]{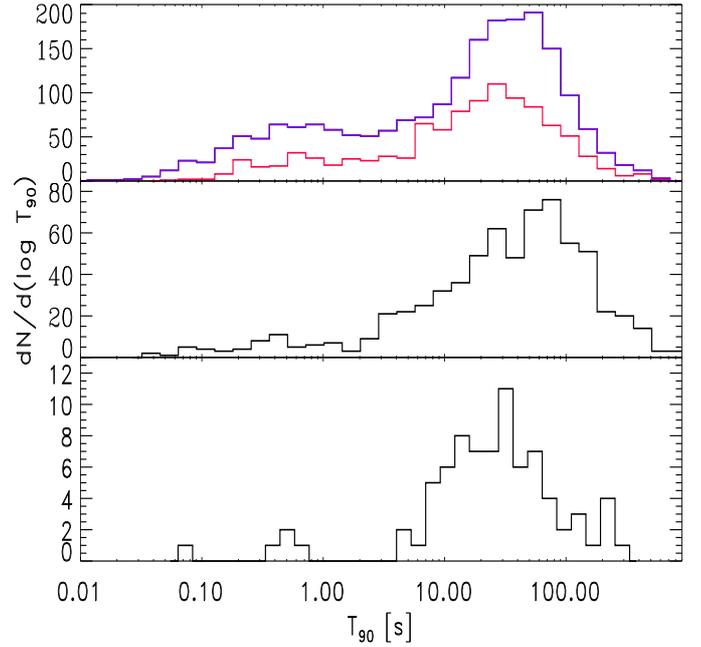}
\caption{Distribution of the duration T$_{90}$. {\it Top:} Distribution of durations derived from BATSE (violet) and {\it Fermi}/GBM (red) light curves in the 50-300 keV band \citep[c.f.][]{kouveliotou93,paciesas12}. {\it Middle:} The T$_{90}$ durations derived using {\it Swift}/BAT instrument on 15-150 keV \citep[c.f.][]{sakamoto11}. {\it Bottom:} Distribution of durations for 20-200 keV light curves obtained from IBIS/ISGRI.}
         \label{fig:t90distr}
\end{figure}

We present an updated version of the currently published {\it INTEGRAL} GRB catalogues in Table \ref{tab:t90} (for the previous versions, see \citealt{vianello09,foley08}). The basic properties of the 28 events observed in the period September 2008 - February 2012 are reported. The information on the counterpart observations are adopted from \INT GRB archive \footnote{http://ibas.iasf-milano.inaf.it}.  The  histogram of the duration T$_{90}$ is shown in Fig. \ref{fig:t90distr}, bottom panel.  For comparison, we also presented the distributions of T$_{90}$ for GRB samples from BATSE\footnote{http://heasarc.gsfc.nasa.gov/W3Browse/cgro/batsegrbsp.html} catalogue, as well as 
{\it Fermi}/GBM\footnote{http://heasarc.gsfc.nasa.gov/W3Browse/fermi/fermigbrst.html} and {\it Swift}/BAT\footnote{http://swift.gsfc.nasa.gov/docs/swift/archive/grb$\_$table.html/} GRBs observed to date. The durations T$_{90}$ of BATSE and {\it Fermi}/GBM bursts were determined using the light curves in 50--300 keV energy band, while {\it Swift} GRB durations were determined using light curves obtained in 15--150 keV energy band. The maximum of the T$_{90}$ distribution for {\it INTEGRAL} GRB sample is at $\sim$ 30 s, which is comparable to the samples obtained by BATSE and {\it Fermi}/GBM. The distinct property of the distribution of T$_{90}$ durations of {\it INTEGRAL} GRBs is the very low number of short gamma-ray bursts with respect to the total number of observed events: only 6\% of the GRBs in the sample have durations $<$ 2 s, compared to 24\%, 17\% and 9\% of short bursts observed by BATSE, {\it Fermi}/GBM, and {\it Swift}/BAT respectively. The paucity of the short events is expected for the imaging instruments, as e.g. the {\it Swift}/BAT, where a minimum number of counts is required to localize an excess in the derived image, making confirmation of real bursts with fewer counts (like the short ones) difficult or impossible even if they are detected by count rate increase algorithms.

\begin{figure*}[!th]
\includegraphics[width=0.5\textwidth]{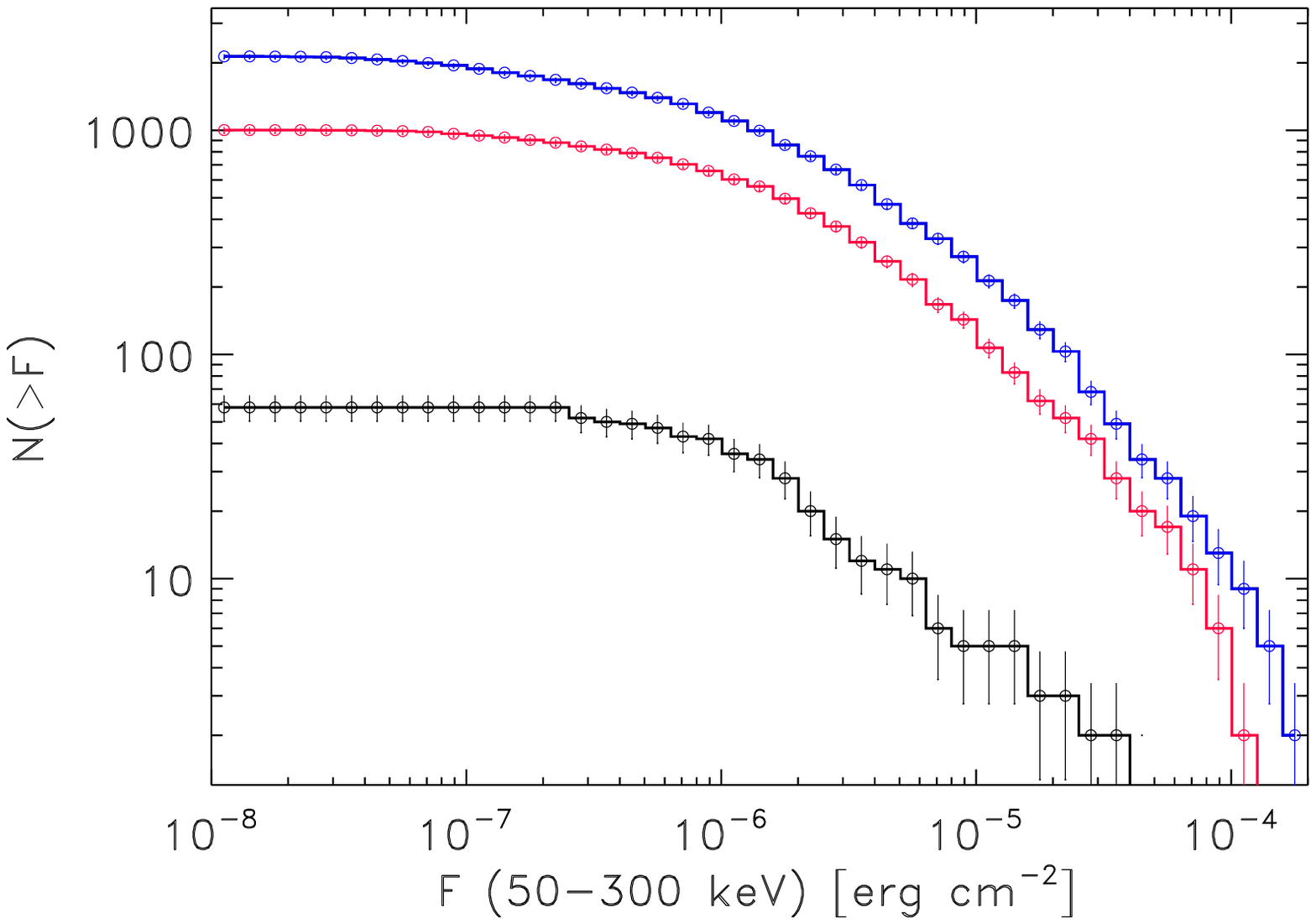}
\includegraphics[width=0.5\textwidth]{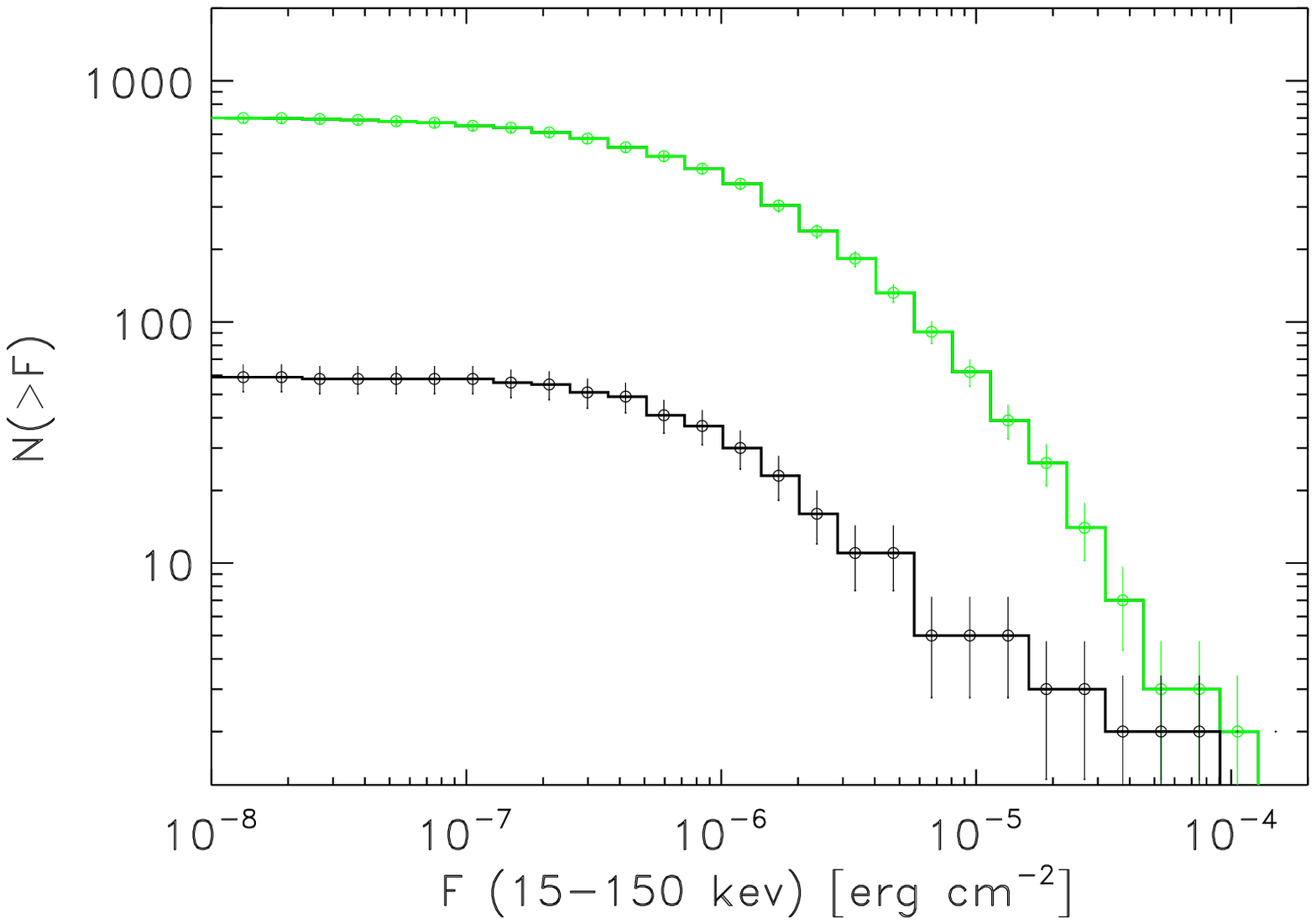}
\caption{Distribution of fluences. {\it Left:} The distributions represent the fluences corresponding to 50--300 keV energy range for the three instruments: BATSE (blue), {\it Fermi GBM} (red), and {\it INTEGRAL} (black). {\it Right:} The distribution of fluences in 15--150 keV energy range for {\it Swift} (green), and {\it INTEGRAL} GRB sample (black).}
\label{fig:fluence}
\end{figure*}

In order to make a comparison with the results from the other GRB missions, we present the cumulative fluence distributions for different instruments in Fig. \ref{fig:fluence}. We calculated the fluences for the \INT\ set of bursts in two different energy bands, 50--300 keV and 15--150 keV, to compare with the data published for the BATSE and {\it Fermi}/GBM GRB samples, and {\it Swift}/BAT GRB sample, respectively. We applied the Kolmogorov--Smirnov (KS) test \citep{press92} and found that the fluence sample of GRBs observed by {\it INTEGRAL} is consistent with the distributions of fluences observed by {\it Swift}/BAT and {\it Fermi}/GBM, with the respective significance probabilities P$_{KS}$ = 0.76 and 0.27. A larger difference is found when \INT\ GRB sample is compared with the distribution corresponding to BATSE sample (P$_{KS}$ = 0.02). Due to the larger sensitivity of the IBIS instrument, one would expected that the larger number of faint GRBs are observed with respect to e.g. {\it Swift}/BAT and {\it Fermi}/GBM missions; however, since {\INT} points for 67\% of its time at low Galactic latitude ($\vert$b$\vert<$20$^\circ$) targets, its sensitivity is affected due to the background induced by the Galactic sources.

\begin{table*}
\caption{{\it INTEGRAL} gamma--ray bursts detected between September 2008 and February 2012. Durations and peak fluxes in the energy band 20-200 keV are reported (see Section \ref{sec:analysis}). We also report the detection of X-ray and optical counterparts.}           
\label{tab:t90}      
\centering                          
\begin{tabular}{c c c c c c c c c c c c c}        
\hline
GRB & t$_{start}$    & R.A.   & Dec.    & Pos.error   & X & O   & T$_{90}$ & Peak flux$^*$ \\ 
    & (UTC)         & (deg)  & (deg)   &  (arcmin)    &  &       &   (s)    & (ph cm$^{-2}$ s$^{-1}$)  \\
\hline
\\                        
081003  & 13:46:01.00  &  262.3764 & 16.5721 & 1.6 & Y &  - &  25$^{+3}_{-3}$   & $<$0.32  \\[0.5ex]
081003B & 20:48:08.00  & 285.0250 & 16.6914 & 1.3& - &  - &  24$^{+6}_{-6}$ & 3.20$^{+0.10}_{-0.10}$  \\[0.5ex]
081016  & 06:51:32.00  &  255.5708 &  -23.3300 & 0.7& Y &  - &  32$^{+5}_{-5}$ &$>$3.30   \\[0.5ex]
081204  & 16:44:56.00  & 349.7750 &  -60.2214 &1.9 & Y &  - &  13$^{+6}_{-6}$ & 0.60$^{+0.40}_{-0.40}$ \\[0.5ex] 
$^a$081226B & 12:13:11.00  & 25.4884  &   -47.4156  & 1.7& - & - &     0.55$^{+0.40}_{-0.40}$ & 0.60$^{+0.50}_{-0.50}$  \\[0.5ex]
090107B & 16:20:38.00  & 284.8075 &  59.5925 & 0.7 & Y & - &   15$^{+3}_{-3}$ &1.50$^{+0.20}_{-0.20}$ \\[0.5ex]
090625B & 13:26:21.00  & 2.2625  &   -65.7817 & 1.5 & Y & - &   10$^{+5}_{-5}$ &2.10$^{+0.10}_{-0.10}$   \\[0.5ex]
090702  & 10:40:35.00  &  175.8883  &     11.5001 & 2.0 & Y & - &   19$^{+8}_{-8}$ & $<$0.20 \\[0.5ex]
090704  & 05:47:50.00  & 208.2042 &  22.7900 & 2.5 & - & - &   76$^{+17}_{-17}$ & 1.30$^{+0.10}_{-0.20}$\\[0.5ex]
090814B & 01:21:14.00  & 64.7750  &   60.5828  & 2.9 & Y & - &   51$^{+12}_{-12}$ & 0.60$^{+0.10}_{-0.10}$\\[0.5ex]
090817  & 00:51:25.00  & 63.9708 &   44.1244  & 2.6 & Y & - &   225$^{+7}_{-7}$ & 2.10$^{+0.10}_{-0.10}$ \\[0.5ex]
091015  & 22:58:53.00  & 306.1292 &  -6.1700  & 2.9 & - & - &   338$^{+77}_{-77}$ & $<$2.37 \\[0.5ex]
091111  & 15:21:14.00  & 137.8125 &  -45.9092  & 2.3 & Y & - &   339$^{+92}_{-92}$ & $<$0.11 \\[0.5ex]
091202  & 23:10:08.00  & 138.8292 &   62.5439 & 2.5 & Y & - &   40$^{+23}_{-23}$ &$<$0.21  \\[0.5ex]
091230  & 06:26:53.00  & 132.8875 &  -53.8925  & 2.5 & Y & Y &   235$^{+36}_{-36}$ & 0.76$^{+0.02}_{-0.03}$\\[0.5ex]
100103A & 17:42:38.00  & 112.3667 &  -34.4825  & 1.1 & Y & - &   35$^{+8}_{-8}$ & 3.40$^{+0.10}_{-0.10}$\\[0.5ex]
100331A & 00:30:23.00  & 261.0625 &  -58.9353  & 2.5 & - & - &   20$^{+6}_{-6}$ & 0.70$^{+0.20}_{-0.20}$\\[0.5ex]
100518A & 11:33:38.00  & 304.7889 & -24.5435 & 1.3  & Y  & Y  &     39$^{+13}_{-13}$ & 0.80$^{+0.20}_{-0.20}$\\[0.5ex]
$^{b}$100703A & 17:43:37.37 &  9.5208&  -25.7097 &  2.6 & - & - &   0.08$^{+0.04}_{-0.04}$  & $<$0.40 \\[0.5ex]
100713A &  14:35:39.00 & 255.2083 &  28.3900  & 2.1 & Y &- &  106$^{+11}_{-11}$ & $<$0.50\\[0.5ex]
100909A & 09:04:04.00 & 73.9500  &  54.6544 & 2.0 & Y&Y &   70$^{+8}_{-8}$ & $<$0.88 \\[0.5ex]
100915B & 05:49:36.00 & 85.3958 &   25.0950  &1.5 & -& - &   6$^{+4}_{-4}$& 0.50$^{+0.10}_{-0.10}$\\[0.5ex]
101112A & 22:10:14.00 &  292.2183  & 39.3589  &0.7 &Y &Y &    24$^{+5}_{-5}$ & $>$1.60\\[0.5ex]
$^{c}$110112B &22:24:54.70 &10.6000  &64.4064 &2.6 &- &- & 0.40$^{+0.15}_{-0.15}$  & 4.60$^{+0.20}_{-0.20}$\\[0.5ex]
110206A &18:07:55.00 &92.3417 &-58.8106 & 1.9 &Y & Y& 35$^{+14}_{-14}$ &1.60$^{+0.20}_{-0.20}$\\[0.5ex]
110708A & 04:43:26.00 & 340.1208 & 53.9597 & 1.2 & Y & - &  79$^{+14}_{-14}$ & 0.80$^{+0.10}_{-0.10}$ \\[0.5ex]
110903A & 02:39:34.00 & 197.0750 & 58.9803 & 0.8 & Y & - &  349$^{+5}_{-5}$ & $>$3.00 \\[0.5ex]
120202A & 21:39:59.00 & 203.5083 & 22.7744 & 1.6 & - & - &  119$^{+6}_{-6}$ & $<$0.20 \\[0.5ex]
\hline                                   
\end{tabular}
\\
\begin{flushleft}
{\small Peak fluxes are calculated for short GRBs on the time interval of $^a$ 0.30 s, $^{b}$ 0.08 s, and $^{c}$ 0.10 s. \\
$^*$The lower limits correspond to GRBs that are heavily affected by telemetry losses.}\\
\end{flushleft}
\end{table*}

\section{Spectral analysis}
\label{sec:spec}
We performed spectral analysis on the whole sample of bursts observed between December 2002 and February 2012. Out of 83 GRBs in the initial sample, we report the results for 59 bursts: for 23 GRBs the data do not provide sufficient signal above background for accurate spectral analysis (three of these events, GRB 021219, GRB 040624 and GRB 050129, were analysed using only IBIS/ISGRI data by \citealt{vianello09}). One gamma-ray burst was observed only during the part of its duration (GRB 080603). The photon spectra were fitted with three models: a single power law, the empirical model for prompt GRB spectra proposed by \citet{band93}, and a cutoff power law model. The counts spectrum, $N(E)$, is given in photons s$^{-1}$ cm$^{-2}$ keV$^{-1}$, and E is in units of keV.  The power-law model is described by N(E) = AE$^\lambda$, and usually characterizes the spectra for which the break energy of the two component spectrum lies outside the instrument energy band, or the spectra for which the signal at high energies is weak and the break energy could not be accurately determined. The other two models we tested allow the determination of the spectral peak energy: the empirical model introduced by \citet{band93},

\begin{eqnarray}
N(E) & = & A (E/100)^\alpha \exp \left( - {E \over E_0} \right); \quad
     {\rm for} \  E \leq \left( \alpha - \beta \right) E_{0}    \nonumber \\ 
    & = & A (E/100)^{\beta} [(\alpha-\beta)(E_0/100)]^{\alpha-\beta} \exp(\beta-\alpha);  
\nonumber \\ 
     &   & {\rm for} \;\; E \geq \left( \alpha - \beta \right) E_{0}  
\end{eqnarray}

\noindent
and power-law model with a high-energy exponential cutoff:
\begin{eqnarray}
N(E) = A E^{\alpha} \exp(-E/E_0).
\end{eqnarray}

\noindent
$E_0$ is the break energy in the Band model; in cutoff power-law model it denotes the {\it e}-folding energy. The cutoff power law model is often used for GRB spectra for which the high energy power law slope $\beta$ in Band model is not well defined due to the low number of high energy photon counts. Using this notation, the peak of the $\nu F_{\nu}$ spectrum for the spectra described by Band or cutoff power-law model is given by $E_{\rm peak}=(\alpha +2)E_{0}$. In Tables \ref{tab:spec} and \ref{tab:spec2} we report the results of the spectral analysis (the best-fit spectral model) of the time-integrated spectra. The fluences in (20--200) keV energy band reported in the table are calculated using the best-fit spectral model. We report also the value of C-statistic and the number of degrees of freedom for the XSPEC spectral analysis.

\begin{table*}
\caption{Results of the spectral fitting: best-fit spectral parameters, the associated 90\% c.l. errors, the value of C-statistic for given d.o.f., and the fluence in 20-200 keV range. The data were fitted with Band model (parameters $\alpha$, $\beta$ and $E_0$), cutoff power law model (parameters $\alpha$ and $E_0$) or a single power law model (parameter $\lambda$). Here $E_0$ is the break energy in the Band model, or the {\it e}-folding energy in case when cutoff power-law model was fitted.}             
\label{tab:spec}      
\centering                          
\begin{tabular}{c c c c c c c c}        
\hline 
GRB & $\alpha$ & $\beta$ & $E_0$ & $\lambda$ & C-STAT/d.o.f. & fluence (20-200 keV)  \\    
    &          &         & [keV] &        &              &  [10$^{-7}$ erg/cm$^2$] \\ [0.5ex]
\hline                      
\\
030227 & -1.03$^{+0.25}_{-0.24}$&-   &97$^{+70}_{-30}$ &-   & 115.8/73  & 6.1$_{-5.9}^{+3.5}$ \\[0.5ex]
030320 &  -                   &-    &-              & -1.39$^{+0.01}_{-0.01}$  &2761.3/866 & 54.2$^{+13.3}_{-11.7}$\\[0.5ex]
030501 & -1.48$^{+0.08}_{-0.08}$&- &184$^{+63}_{-38}$ &- & 690.2/249 & 17.2$_{-3.1}^{+1.6}$\\[0.5ex]
030529 &-                    &-     &-&  -1.61$^{+0.10}_{-0.10}$  &151.3/74 & $<$ 5.3\\[0.5ex]
031203 & -    &- &- &  -1.51$^{+0.03}_{-0.03}$  &477.4/162 & 10.6$_{-3.0}^{+2.7}$\\[0.5ex]
040106 &  -1.27$^{+0.23}_{-0.18}$&- &$>$135 &-&265.5/117 & 9.5$_{-9.1}^{+2.3}$ \\[0.5ex]
040223 &   -  &- &- &  -1.73$^{+0.06}_{-0.07}$  &126.1/75 & 27.2$^{+0.8}_{1.9}$\\[0.5ex]
040323 &-0.50$^{+0.09}_{-0.09}$ &-&174$^{+44}_{-31}$&- &538.9/381 & 20.6$_{-2.9}^{+2.3}$\\[0.5ex]
040403  &-0.75$^{+0.38}_{-0.35}$ &-&68$^{+56}_{-23}$ &-&232.8/161 & 4.0$_{-3.7}^{+1.6}$\\[0.5ex]
040422 &-0.33$^{+0.30}_{-0.28}$ &-&27$^{+5}_{-4}$ &-&272.4/161 &4.9$_{-3.6}^{+1.0}$ \\[0.5ex]
040730 &  -   &- &- &-1.25$^{+0.07}_{-0.07}$  & 166.4/118 & 6.3$_{-3.3}^{+4.4}$ \\[0.5ex]
040812 & -   &- &- &-2.10$^{+0.14}_{-0.15}$ &94.1/74 & $<$ 6.9\\[0.5ex]
040827 &  -0.34$^{+0.21}_{-0.20}$      &- &54$^{+12}_{-9}$  & -&668.7/293 & 11.1$_{-4.0}^{+2.8}$\\[0.5ex]
041218 &   -  & -&- &-1.48$^{+0.01}_{-0.01}$ &768.5/250 &58.2$_{-3.7}^{+3.5}$ \\[0.5ex]
041219 &  -1.48$^{+0.14}_{-0.11}$    &-2.01$^{+0.05}_{-0.08}$ &301$^{+170}_{-105}$  &- &280.7/304 &867.3$_{-128.9}^{+5.4}$ \\[0.5ex]
050223 &   -  &- &- &-1.44$^{+0.06}_{-0.06}$  &342.5/161 & $<$ 15.7 \\[0.5ex]
050502 &    -1.07$^{+0.13}_{-0.13}$  & - & 205$^{+132}_{-63}$ &- &324.0/293 &13.9$_{-4.0}^{+1.1}$ \\[0.5ex]
050504 &   -  & -&- &-1.01$^{+0.05}_{-0.04}$ &122.0/74 & 10.0$_{-4.5}^{+4.1}$\\[0.5ex]
050520 &   -  &- &- &-1.45$^{+0.04}_{-0.03}$  & 119.4/74& 16.6$_{-5.0}^{+4.9}$\\[0.5ex]
050525 &   -1.09$^{+0.04}_{-0.04}$     &-  &131$^{+12}_{-10}$  &- &2511.6/733 & 153.9$_{-8.4}^{+5.7}$\\[0.5ex]
050626 &   -  & -&- & -1.11$^{+0.13}_{-0.13}$  &33.3/31 & 6.3$_{-1.0}^{+0.4}$\\[0.5ex]
050714 &  -   &- &- & -1.63$^{+0.10}_{-0.11}$  & 99.3/74&$<$ 4.5\\[0.5ex]
050918 &  -   &- &- & -1.50$^{+0.02}_{-0.02}$  &1476.5/866 & 30.2$_{-9.0}^{+10.5}$\\[0.5ex]
051105B &  -   &- &- & -1.57$^{+0.11}_{-0.12}$  &290.5/250 & 2.8$_{-2.0}^{+1.5}$\\[0.5ex]
051211B &  -   &- & -& -1.38$^{+0.04}_{-0.04}$& 304.7/162&  16.1$_{-3.3}^{+4.6}$\\[0.5ex]
060114 &  -   &- & -& -0.80$^{+0.07}_{-0.08}$ &89.0/31 & 16.0$_{-1.5}^{+0.7}$\\[0.5ex]
060204 &  -   & -&- & -1.13$^{+0.11}_{-0.11}$   &217.2/162 & 4.8$_{-3.3}^{+2.4}$\\[0.5ex]
060428C &  -0.90$^{+0.14}_{-0.12}$       &-1.88$^{+0.14}_{-0.29}$ & 108$^{+34}_{-26}$&- &179.0/116 & 18.6$_{-3.9}^{+2.2}$\\[0.5ex]
060901 & -1.11$^{+0.06}_{-0.05}$    &- &265$^{+71}_{-49}$ &-  &540.7/293 &62.2$_{-5.9}^{+3.5}$ \\[0.5ex]
060912B &  -   &- &- &-1.34$^{+0.08}_{-0.08}$ &174.5/162 &12.0$_{-5.1}^{+5.8}$ \\[0.5ex]
061025 & -0.53$^{+0.21}_{-0.20}$ &- &87$^{+35}_{-20}$&-  &227.8/117 & 10.1$_{-4.8}^{+1.3}$\\[0.5ex]
061122 &    -1.30$^{+0.05}_{-0.05}$    &  -&263$^{+35}_{-30}$ & -&685.3/513 &155.1$_{-5.3}^{+3.4}$ \\[0.5ex]
070309 & 0.43$^{+0.78}_{-0.63}$     &- &45$^{+39}_{-16}$ &- &174.5/73 & $<$ 12.6 \\[0.5ex]
070311 &   -0.84$^{+0.08}_{-0.15}$   &- &  266$^{+199}_{-88}$& - &449.9/205 &23.6$_{-5.3}^{+1.7}$ \\[0.5ex]
070925 &  -1.06$^{+0.09}_{-0.08}$  & -&317$^{+135}_{-80}$& -  &497.6/337 &36.1$_{-3.4}^{+1.7}$ \\[0.5ex]
071109 &  -   & -& -&  -1.31$^{+0.08}_{-0.08}$   &166.5/118 & 3.6$_{-3.5}^{+4.0}$\\[0.5ex]
080613 &   -1.00$^{+0.17}_{-0.12}$  &- &$>$202 &  -   &187.0/117 & 12.3$_{-5.9}^{+1.7}$ \\[0.5ex]
080723B &   -1.01$^{+0.02}_{-0.02}$    & - & 326$^{+30}_{-26}$ &- &535.2/293 &396.4$_{-6.7}^{+6.7}$ \\[0.5ex]
080922 &  -   &- &- &-1.72$^{+0.03}_{-0.03}$ &274.8/162 &17.3$_{-6.5}^{+6.9}$ \\[0.5ex]
081003B &  -1.31$^{+0.07}_{-0.04}$   &- &$>$435 & -&598.1/381 & 26.2$_{-24.5}^{+2.0}$\\[0.5ex]
081016 &  -1.09$^{+0.12}_{-0.12}$ &-& 135$^{+48}_{-29}$ &-&509.8/425 &22.0$_{-4.5}^{+1.4}$ \\ [0.5ex]
081204 &-1.34$^{+0.27}_{-0.25}$ &-&110$^{+139}_{-42}$ &- &504.0/249 &5.1$_{-4.8}^{+5.1}$ \\ [0.5ex] 
090107B &-1.20$^{+0.16}_{-0.15}$ &-&217$^{+265}_{-81}$ & - &304.1/205 &12.4$_{-4.6}^{+1.3}$ \\[0.5ex]
090625B & -0.47$^{+0.13}_{-0.13}$  &-   &104$^{+27}_{-18}$ & - &405.5/205 &12.4$_{-2.0}^{+1.2}$ \\[0.5ex]
090702 & -1.19$^{+0.54}_{-0.72}$  &- &46$^{+165}_{-25}$ &-& 247.1/117 &$<$ 2.1 \\ [0.5ex]
\hline                                   
\end{tabular}
\end{table*}

\begin{table*}
\caption{Results of the spectral fitting. (continued)}             
\label{tab:spec2}      
\centering                          
\begin{tabular}{c c c c c c c c}        

\hline
GRB & $\alpha$ & $\beta$ & $E_0$ & $\lambda$  & C-STAT/d.o.f.& fluence (20-200 keV)  \\    
    &  & & [keV] & &  &  [10$^{-7}$ erg/cm$^2$] \\ [0.5ex]
\hline                   
\\
090704 &  -1.19$^{+0.06}_{-0.06}$&- &447$^{+276}_{-135}$ &-&1516.9/557 & 54.1$_{-8.0}^{+4.9}$\\[0.5ex] 
090814B &  -   & -&- &-0.94$^{+0.04}_{-0.04}$  &470.2/250 &15.1$_{-2.4}^{+2.3}$ \\[0.5ex]
090817 &  -   &- & -   &-1.39$^{+0.04}_{-0.05}$  &110.9/74 & 18.7$^{+10.9}_{-9.8}$\\[0.5ex]
091015  &  -   &- &- &-1.36$^{+0.07}_{-0.07}$ & 280.8/118&$<$ 30.2 \\[0.5ex]
091111  &   -  &- &- &-0.99$^{+0.07}_{-0.09}$  & 286.4/250&$<$ 12.2\\[0.5ex]
091202  & -    &- &- &-1.07$^{+0.09}_{-0.09}$  & 170.7/118& $<$ 4.2\\[0.5ex]
100103A &  -0.85$^{+0.06}_{-0.06}$   &- &222$^{+48}_{-35}$  &- &731.1/381 & 52.5$_{-4.0}^{+2.1}$\\[0.5ex]
100518A &- & -& -&-1.28$^{+0.05}_{-0.05}$ &410.9/162  & 5.2$_{3.8}^{+4.4}$ \\ [0.5ex]
100713A &-  &- &- &-1.44$^{+0.09}_{-0.09}$  &381.0/206 &$<$ 4.5 \\[0.5ex]
100909A &-0.38$^{+0.24}_{-0.21}$   &- & 181$^{+192}_{-69}$ &- &218.2/73 & $<$ 19.3\\[0.5ex]
101112A &-0.93$^{+0.14}_{-0.14}$   &- &251$^{+279}_{-91}$ &-  &141.0/73 & 21.1$_{-7.4}^{+4.4}$\\[0.5ex]
110708A &-0.90$^{+0.11}_{-0.11}$ & - &143$^{+48}_{-30}$ & - & 796.5/469 & 24.8$_{-4.6}^{+1.9}$\\[0.5ex]
110903A &-0.73$^{+0.04}_{-0.04}$ & - & 484$^{+165}_{-102}$ & - & 1490.7/469 &148.0$_{-17.5}^{+11.9}$ \\[0.5ex]
120202A &-1.09$^{+0.25}_{-0.17}$ & - & $>$130 & - &  455.7/425 &8.0$_{-7.7}^{+2.1}$ \\[0.5ex] 
\hline                                   
\end{tabular}
\end{table*}

\begin{figure*}[!t]
\begin{center}  
 \begin{tabular}{cc}
\centering
\includegraphics[width=0.5\textwidth]{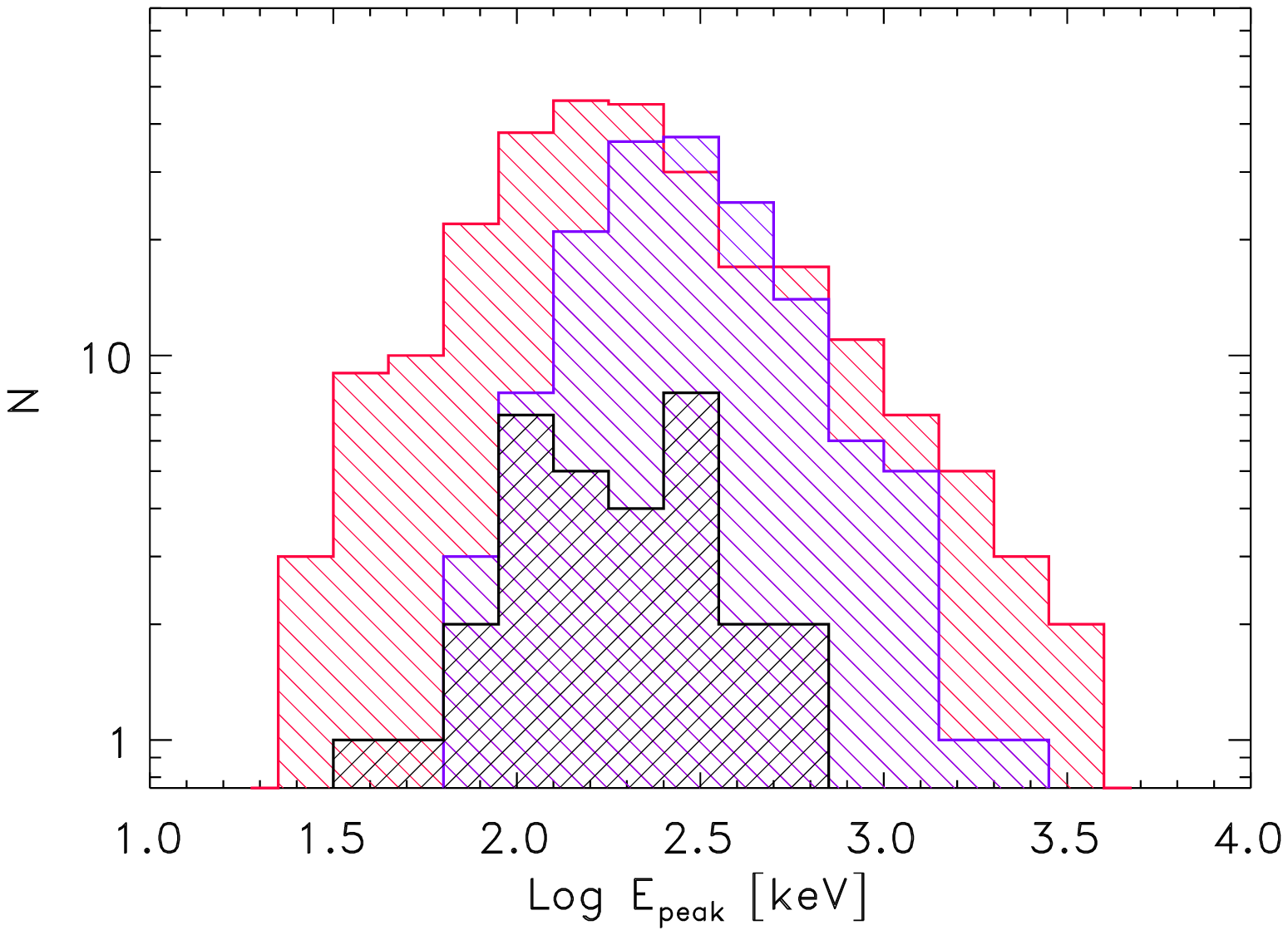}&
\includegraphics[width=0.5\textwidth]{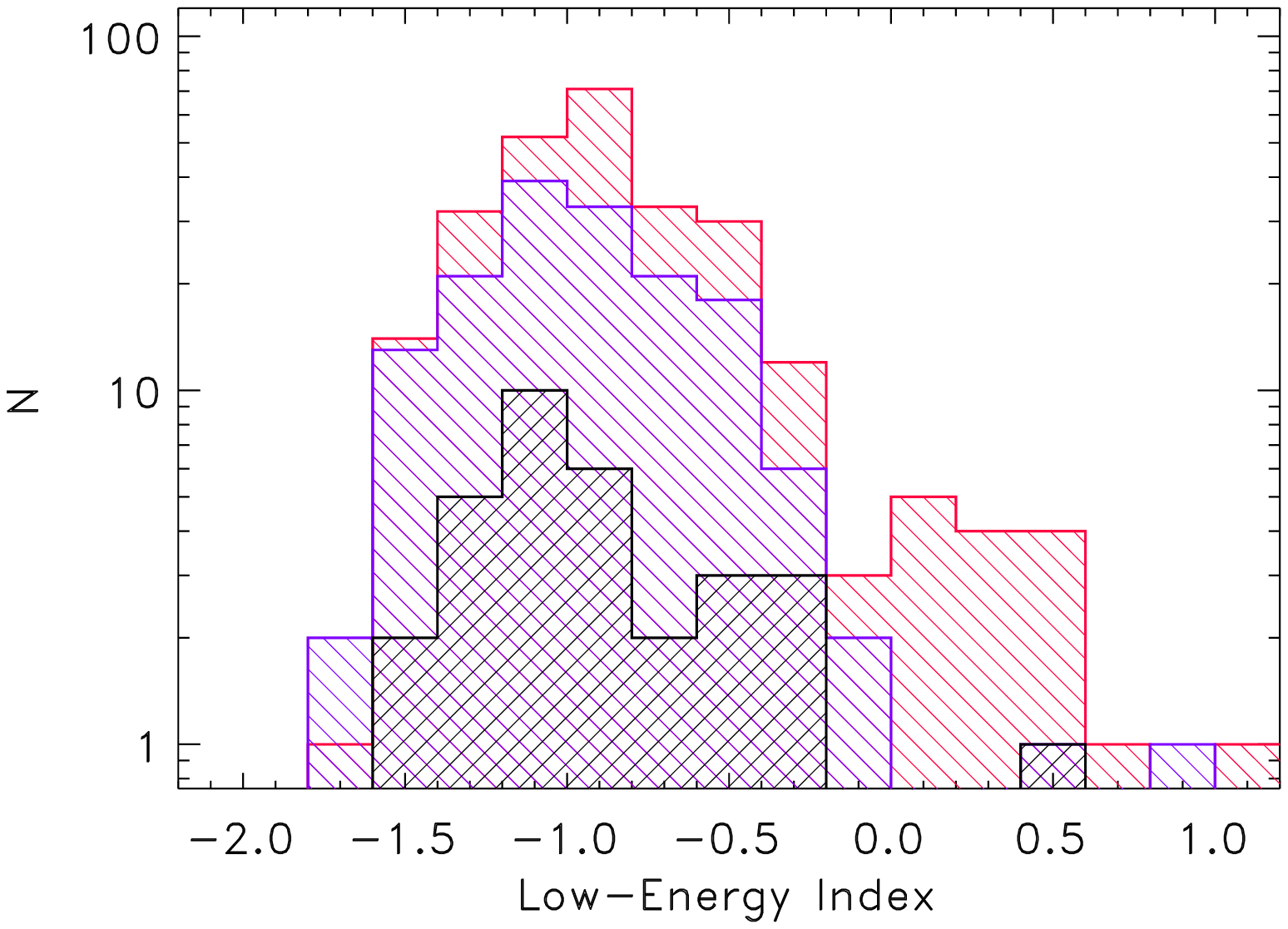}\\
  \end{tabular}
\caption{Distribution of the spectral peak energies (left) and the low energy spectral power law indices (right). The sample of {\it INTEGRAL} GRBs is shown in black; BATSE results (violet) and {\it Fermi}/GBM results (red) of the time-integrated spectral analysis were used for the comparison. Only long events were selected, fitted with the Band or cut-off power law model, and having the fluence in the same range as {\it INTEGRAL} GRBs.}
        \label{fig:peak}
\end{center}
\end{figure*}

{\it Distribution of E$_{peak}$}. We show the histogram of the observed values of the spectral peak energy in Fig. \ref{fig:peak}, left panel. It contains the results obtained by fitting the Band or the cut off power law model to the time-integrated spectra of the \INT\ bursts. There were 30 GRBs that were fitted with the cut off power law model and 2 GRBs fitted with Band model in our sample. We compared these results with the results obtained by {\it Fermi}/GBM and BATSE detectors. It was shown by various authors that the observed E$_{peak}$ correlates with the burst brightness (e.g. \citealt{mallozzi95,lloyd00,kaneko06,nava08}); in order to account for the possible biases in the distribution of the spectral parameters, we made a comparison of the GRBs within the same fluence range (see also \citealt{nava11} for the comparison between the spectral properties of {\it Fermi}/GBM and BATSE gamma-ray bursts). We selected from the {\it Fermi}/GBM$^5$ and BATSE$^4$ databases the results of the analysis obtained for: (i) GRBs within same fluence range as \INT\ GRBs (i.e. fluence in 50-300 keV energy range $<$ 8.7 $\times$ 10$^{-5}$ erg cm$^{-2}$). The lower fluence limit is approximately the same for all three samples ($\sim$10$^{-8}$ erg cm$^{-2}$). Only the condition on the fluence limit was imposed since the peak fluxes were determined on 20--200 keV energy band for {\it INTEGRAL} GRBs, while the {\it Fermi}/GBM and BATSE databases contain the values of peak fluxes determined on 50--300 keV; 
(ii) GRBs with durations of T$_{90} >$ 2 s; (iii) GRBs for which the model that best fitted the time-integrated spectrum was the Band or the cutoff power-law model. We did not apply any additional condition based on the quality of the spectral fit (c.f. \citealt{goldstein12}, \citealt{kaneko06}). The histograms for {\it Fermi}/GBM bursts (red line) and BATSE bursts (violet line) are shown in Fig. \ref{fig:peak}.  We used the Kolmogorov-Smirnov test in order to establish the probability whether the distribution of the spectral parameters of the

\begin{figure}
\includegraphics[width=0.5\textwidth]{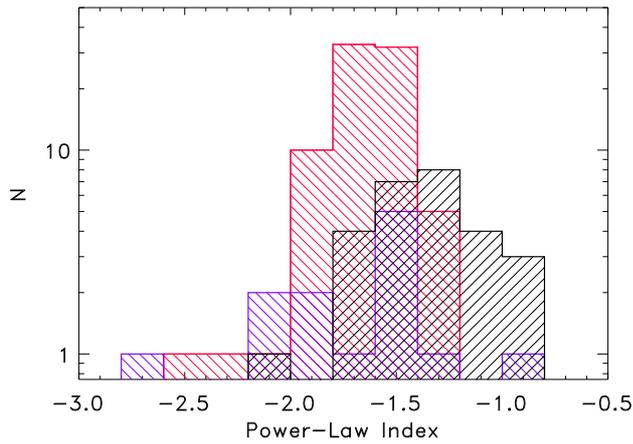}
\caption{Distribution of the power law indices for the sub-sample of {\it INTEGRAL} GRBs that were fitted with the single power law model on the energy range 20-1000 keV (black line). BATSE results (violet) and {\it Fermi GBM} results (red) are shown for the gamma-ray bursts for which the best spectral model was a single power law. The results for the analysis of the time-integrated spectra are shown, using GRBs within the same fluence range as {\it INTEGRAL} GRBs.}
         \label{fig:pl}
\end{figure}

\begin{figure}
\includegraphics[width=0.5\textwidth]{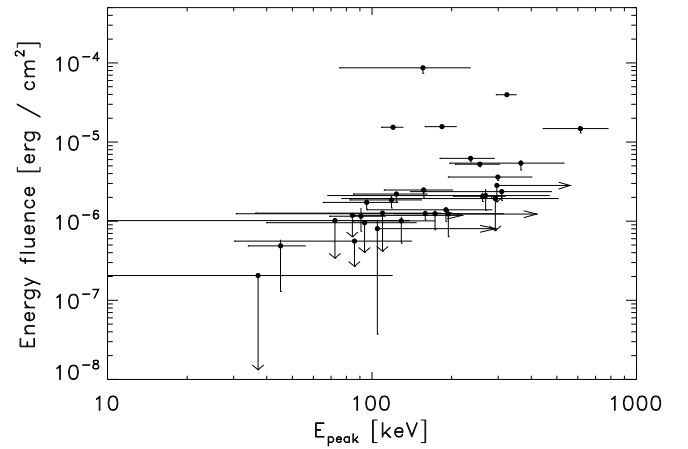}
\caption{Correlations between spectral parameters. Energy fluence in 20-200 keV vs. spectral peak energy E$_{peak}$.}
\label{fig:epfl}
\end{figure}

\noindent
{\it INTEGRAL} GRBs can be derived from the same parent distribution as {\it Fermi}/GBM or BATSE bursts. For the distribution of spectral peak energies, we found that our distribution is consistent with the distribution of the spectral peak energies of GRBs observed by {\it Fermi}/GBM (KS probability = 0.55), and not consistent with the distribution of BATSE GRBs (KS probability = 6 $\times$ 10$^{-3}$) in a given fluence range. 

{\it Distribution of $\alpha$.} The distribution of the low energy spectral slopes for {\it INTEGRAL} GRBs is shown in Fig. \ref{fig:peak}, right panel. We compared this distribution with the parameters of the {\it Fermi}/GBM and BATSE GRB samples. The selection of GRBs from {\it Fermi}/GBM and BATSE samples was done in the same way as it was in the case of E$_{peak}$ distribution. The distribution of the low energy power law slopes obtained for ISGRI/SPI GRBs is consistent with both, {\it Fermi}/GBM (KS probability = 0.23) and BATSE (KS probability = 0.92) GRB samples. 

{\it Distribution of $\lambda$.} In the {\it INTEGRAL} sample there were 27 GRBs for which the model that best fitted the data was a single power law with the slope $\lambda$ (see Table \ref{tab:spec}). Fig. \ref{fig:pl} shows the distribution of $\lambda$ for our sample. For the comparison we plotted the results obtained for the BATSE and {\it Fermi}/GBM sample of GRBs for which the best fitted model was a single power law. We selected only the long bursts within the same fluences range as the {\it INTEGRAL} GRBs. We find that the distribution obtained for the {\it INTEGRAL} GRBs is not consistent with the distribution corresponding to {\it Fermi}/GBM population (KS probability = 5 $\times$ 10$^{-6}$), and is consistent with the distribution of BATSE GRBs (KS probability = 0.05).

\begin{figure*}[!th]
\begin{center}  
 \begin{tabular}{cc}
\centering
\includegraphics[width=0.5\textwidth]{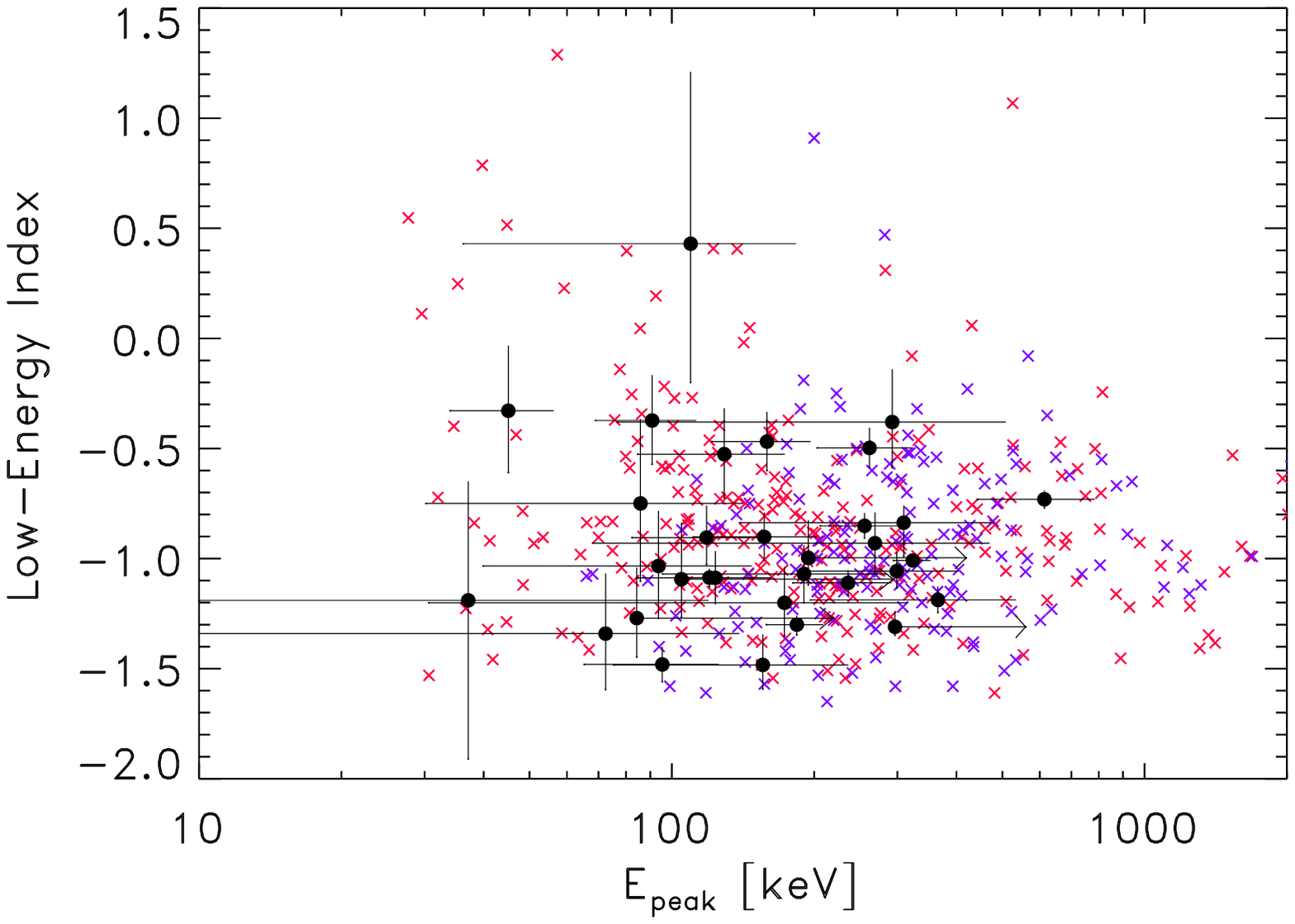}&
\includegraphics[width=0.5\textwidth]{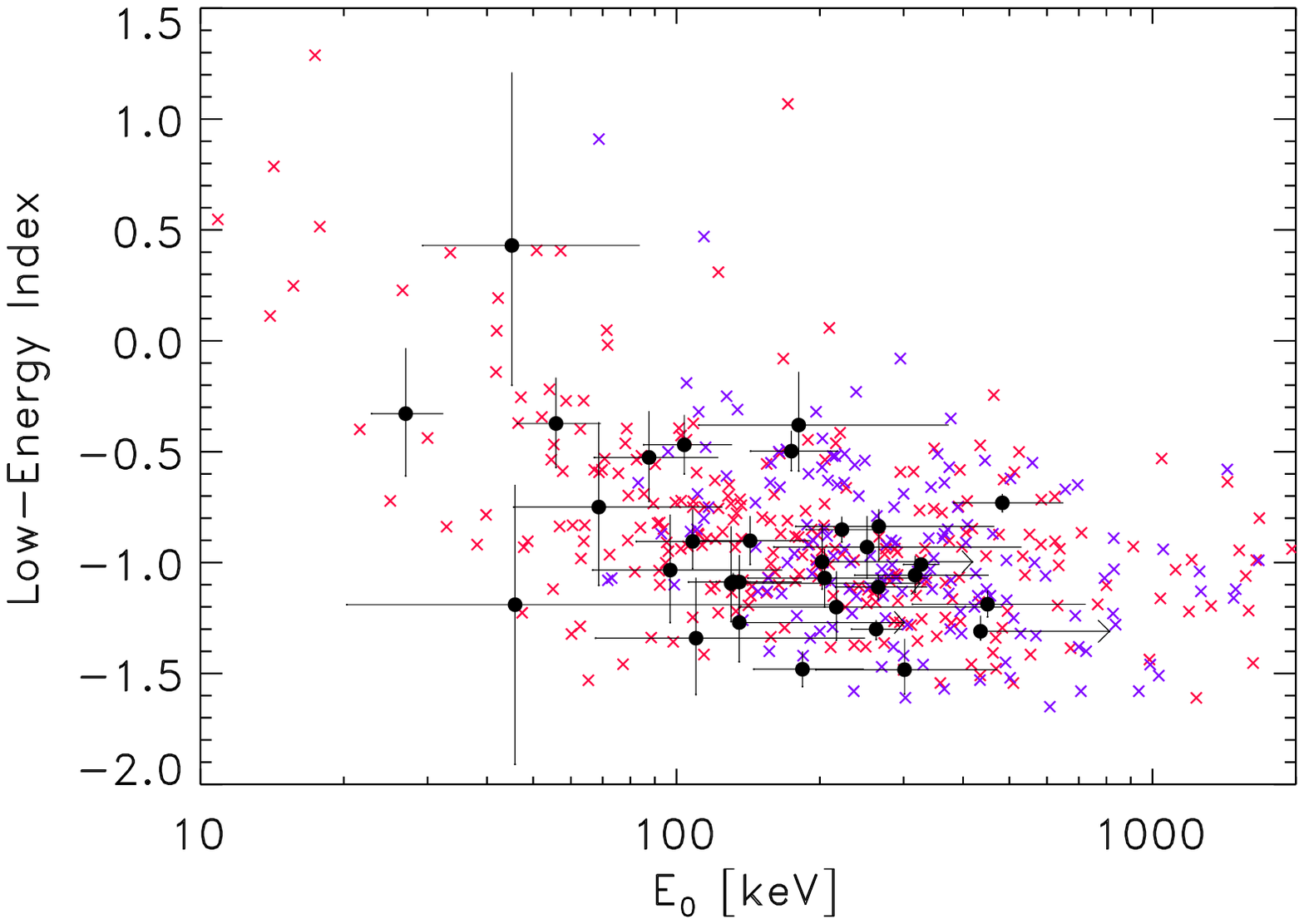}\\
  \end{tabular}
\caption{Correlations between spectral parameters. Low energy spectral index $\alpha$ vs. spectral peak energy E$_{peak}$ ({\it left}) and low energy spectral index $\alpha$ vs. break energy E$_0$ ({\it right}). For the reference, we also show the parameters of the time-integrated spectral analysis for {\it Fermi} (red) and BATSE (violet) GRBs.}
         \label{fig:corr}
\end{center}
\end{figure*}

{\it Correlations among spectral parameters}. The empirical correlations among time-resolved spectral parameters were examined for BATSE and {\it Fermi}/GBM samples \citep{crider97,preece98,lloyd02,kaneko06,goldstein12}. The most significant correlation is found between E$_{peak}$ and low energy spectral index for the time resolved spectra of individual bursts. The correlations between the time-integrated parameters E$_{peak}$ and $\alpha, \beta$ and energy or photon flux/fluence were also investigated \citep{kaneko06,goldstein12}. We show in Figs. \ref{fig:epfl} and \ref{fig:corr} energy fluence in 20--200 keV vs. E$_{peak}$ and the scatter plots $\alpha$ vs. E$_{peak}$, $\alpha$ vs. E$_{0}$. For reference, we show the parameters resulting from the spectral analysis of time-integrated spectra for {\it Fermi}/GBM and BATSE GRB samples.
The general trend is that lower measured spectral peak energies (close to the lower end of the instrument energy band) increase the uncertainty of the low-energy power law index (c.f. \citealt{goldstein12} for the sample of {\it Fermi}/GBM GRBs). We calculated the Spearman rank-order correlation coefficient (r$_s$) and the corresponding significance probability P$_{rs}$ to test the existence of correlations among pairs of the time integrated parameters, low energy spectral index -  E$_{peak}$ and low energy spectral index -  E$_{0}$.
We found no significant correlation in the first case, while for the latter one there exists a marginal negative correlation (r$_s$ = --0.44) with the associated significance probability P$_{rs}$ = 1.15 $\times$ 10$^{-2}$. Among the energy fluence - E$_{peak}$, see Fig. \ref{fig:epfl}, we found a weak positive correlation (r$_s$ = 0.50) with the associated significance probability P$_{rs}$ = 1.88 $\times$ 10$^{-2}$. \citet{kaneko06} examining the correlations among time-integrated parameters also found only one significant correlation, namely between E$_{peak}$ and total energy fluence.

\section{Summary}

We have presented a spectral catalogue of the GRBs observed by the \INT instruments in the period December 2002 - February 2012. We developed a new spectral extraction method especially suited for short transients where total number of counts is small.  We are nevertheless able to probe the high spectral end of the \INT instruments' energy range thanks to the use of the Cash statistic. This new method has been applied in a coherent way to the already previously published GRBs, as well as to the unpublished ones. It allowed us to measure the time integrated GRB peak energy in about 54\% of the GRBs of our sample, while for the most complete \INT GRB sample published to date \citep{vianello09} this fraction was just 16\% (for BAT data, this fraction corresponds to 17\%, \citealt{sakamoto11}). This has allowed us for the first time to fully compare the \INT sample to previous and current GRB dedicated experiments' results in the spectral and temporal domain.

\begin{table}
\caption{The median parameter values and the dispersion (quartile) of the distributions obtained for time-integrated spectra fitted with Band or cutoff power-law model.}           
\label{tab:res}      
\centering                          
\begin{tabular}{c c c c}        
\hline
Low energy index & E$_{0}$ [keV]    & E$_{peak}$ [keV]  & $\lambda$  \\ 
\hline
\\ 
 --1.01$^{+0.28}_{-0.18}$&205$^{+97}_{-97}$ & 184$^{+110}_{-65}$& --1.39$^{+0.26}_{-0.12}$ \\                       
\\
\hline                                   
\end{tabular}
\end{table}

Our temporal analysis showed that the T$_{90}$ duration distribution of \INT GRBs is comparable to the one of BAT bursts, showing the paucity of the short GRBs with respect to the GRB samples detected by {\it Fermi}/GBM and BATSE.
The maximum of the distribution of T$_{90}$ durations is at $\sim$ 30 s, which on the other hand makes it similar to the {\it Fermi}/GBM sample. The reason for that lies in the triggering time scales of the two instruments: in case of IBAS, {\it Fermi}/GBM, and BATSE it is of the order of tens of seconds, while {\it Swift}/BAT  triggering time scales can be as long as tens of minutes.     

Concerning the GRB fluence distribution of our sample, we found it statistically compatible with the {\it Swift}/BAT and {\it Fermi}/GBM ones. While the IBAS system is expected to be intrinsically more sensitive than {\it Swift}/BAT or {\it Fermi}/GBM (see Fig. \ref{fig:sensi}), the fact that \INT spends most of its observing time pointing Galactic sources implies a diminished sensitivity due to the increased background induced by these sources. 

In Table \ref{tab:res} we report the median spectral parameter values and the dispersions for the distributions obtained for \INT GRB time-integrated spectra. The peak energy values we could determine are compatible with the ones obtained by the {\it Fermi}/GBM experiment, and not with the ones measured by BATSE, being systematically softer. This can be explained by the similar triggering threshold of the two former instruments (15 keV vs. 8 keV), which are both significantly lower than the nominal BATSE low energy threshold of 50 keV. The median of the peak energy distribution is at $\sim$180 keV, with a dispersion of $\sim$100 keV. The slopes of the low energy photon spectra are found to be consistent with both samples, {\it Fermi}/GBM and BATSE, having the median of the distribution at $\alpha$=--1 and a spread $\lesssim$0.3. When a single power law was fitted to the spectra, we found that the distribution of power law indices has its median at $\lambda$=--1.4, and a spread $\lesssim$0.3. \INT GRBs that are fitted with a single power-law are therefore harder when compared with {\it Fermi}/GBM sample, and consistent with the BATSE GRB sample. The correlations among the spectral properties (e.g. low energy spectral index vs. spectral peak energy)  were investigated for the time-resolved spectra of the individual GRBs, and were not tested in this work due to the insufficient count number. We confirm that the analogous correlation among the time-integrated spectral properties does not hold for \INT GRB sample, as it was also found for e.g. BATSE data by \citet{kaneko06}. Weak correlations were found for low energy spectral index $\alpha$ vs. break energy E$_0$, and energy fluence vs. the observed spectral peak energy. 

The GRB catalog we presented contains a limited number of events with respect to other missions' databases. Our results allow however an important insight in the possible instrumental biases in spectral and temporal parameters distributions, and also provide the spectral analysis for a sample of faint GRBs with good statistics.

\begin{acknowledgements}
The authors thank Jochen Greiner for careful reading of the manuscript, and valuable comments on this work. The authors thank Thomas Maccarone, Patrick Sizun and Fabio Mattana for discussions on data analysis. ZB acknowledges the French Space Agency (CNES) for financial support. ISGRI has been realized and maintained in flight by CEA-Saclay/Irfu with the support of CNES. Based on observations with INTEGRAL, an ESA project with instruments and science data centre funded by ESA member states (especially the PI countries: Denmark, France, Germany, Italy, Switzerland, Spain), Czech Republic and Poland, and with the participation of Russia and the USA. 
\end{acknowledgements}

\bibliographystyle{aa} 
\bibliography{bosnjak_integral}

\newpage
\begin{appendix}
\begin{figure*}[!t]
   \begin{center}
\section{IBIS/ISGRI light curves of {\it INTEGRAL} GRBs}

   \begin{tabular}{cc}
   \centering
\includegraphics[width=0.5\textwidth]{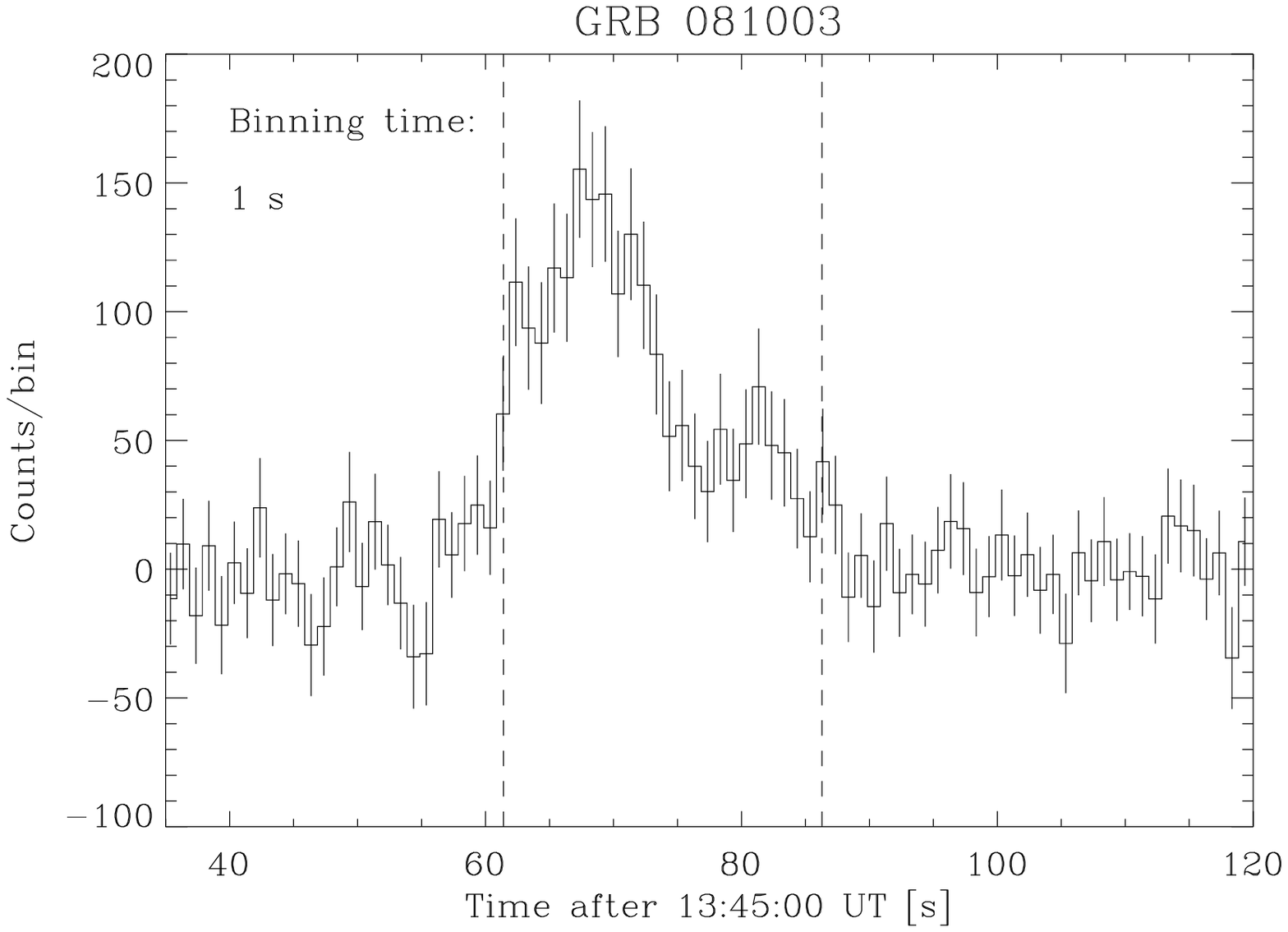}&

 \includegraphics[width=0.5\textwidth]{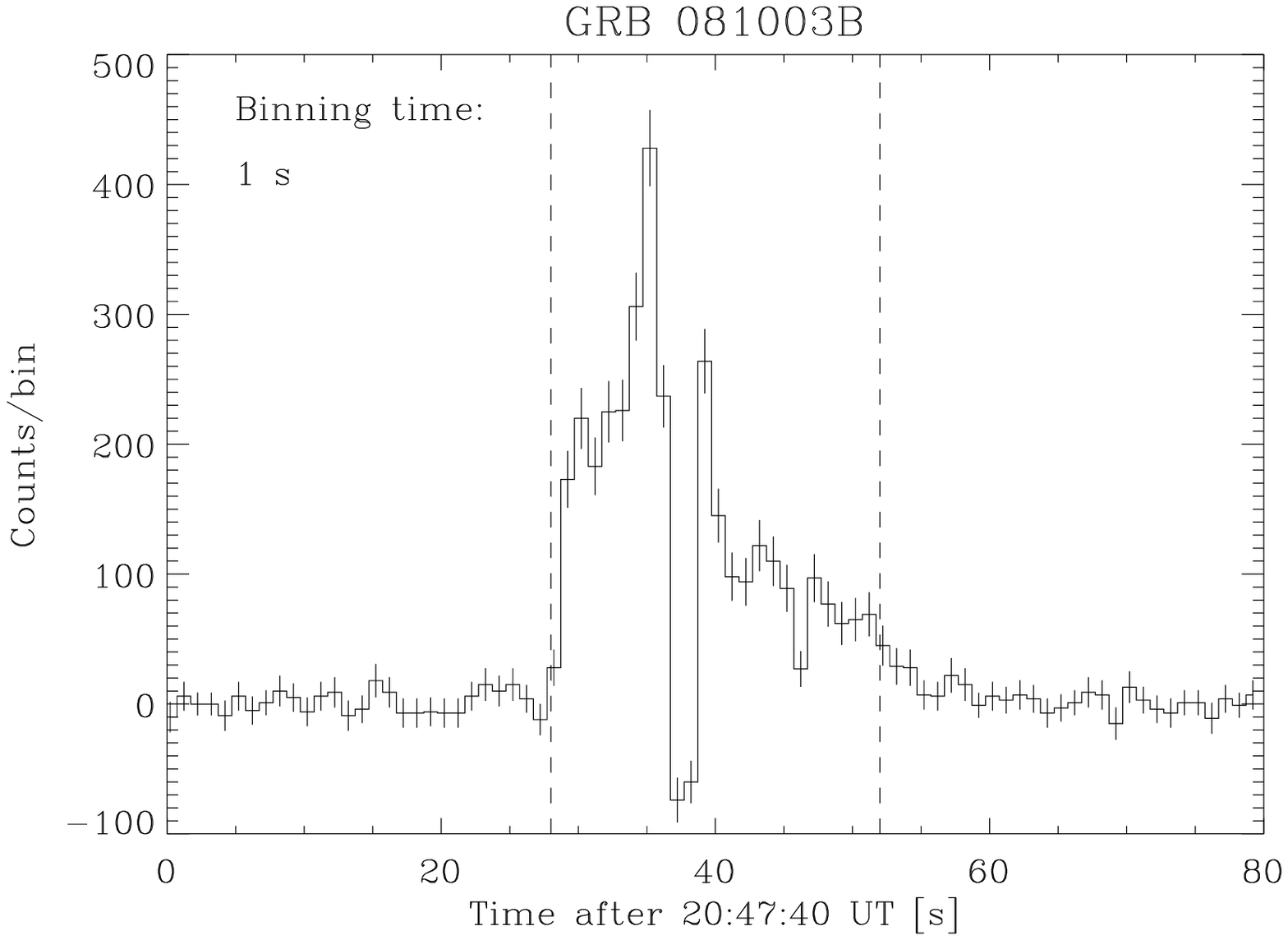}\\
   \includegraphics[width=0.5\textwidth]{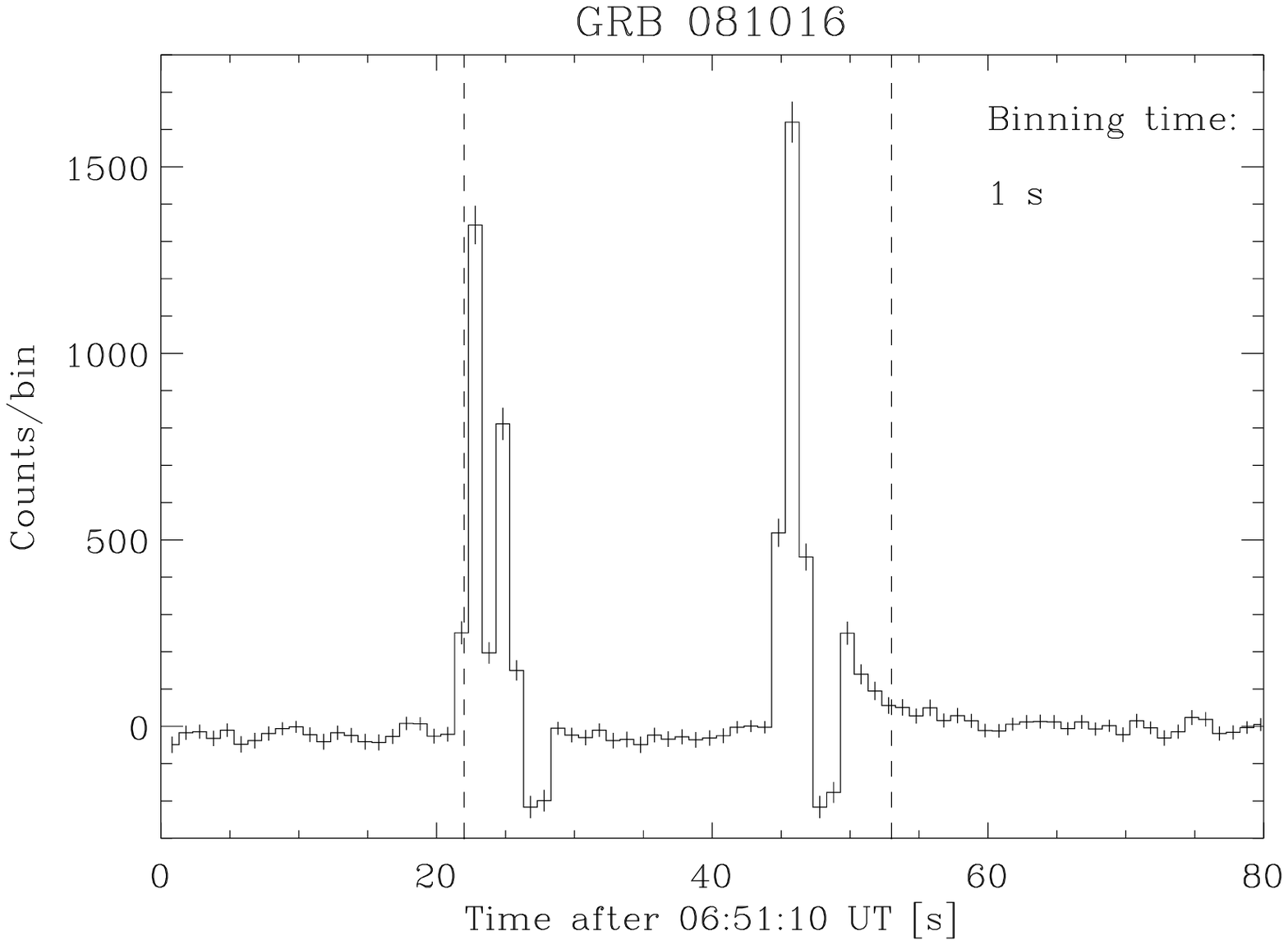}&

   \includegraphics[width=0.5\textwidth]{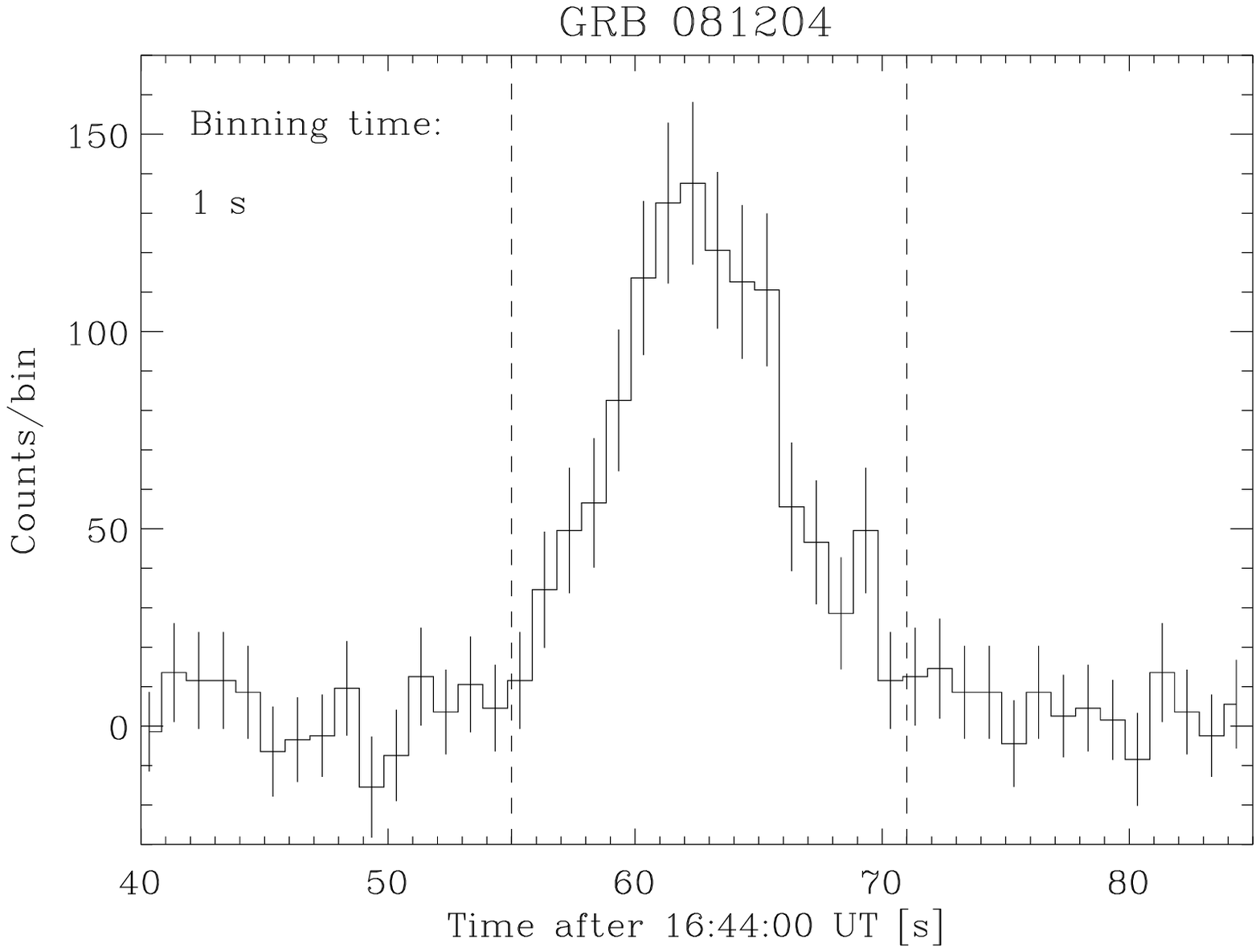}\\
   \includegraphics[width=0.5\textwidth]{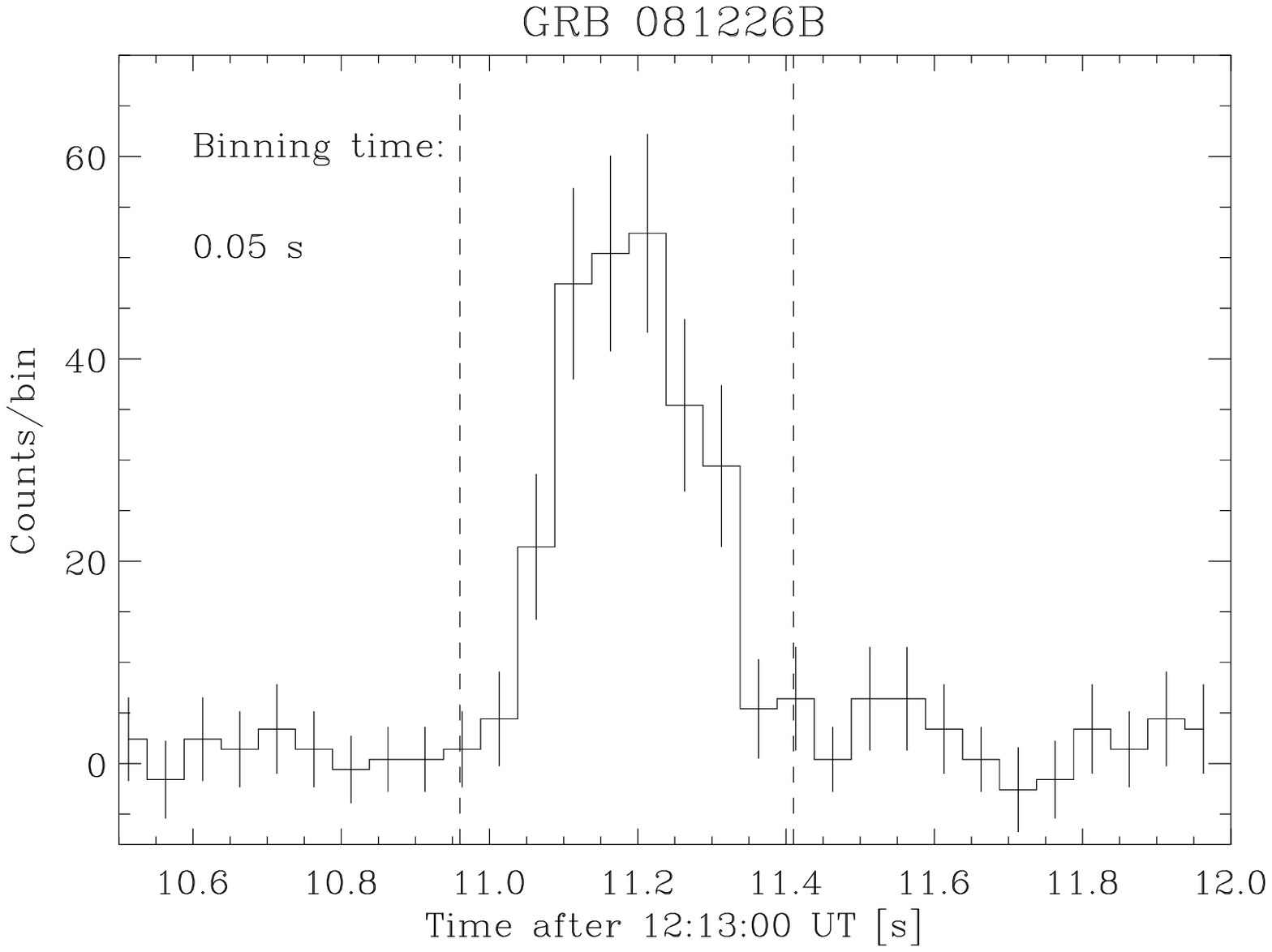}&
   \includegraphics[width=0.5\textwidth]{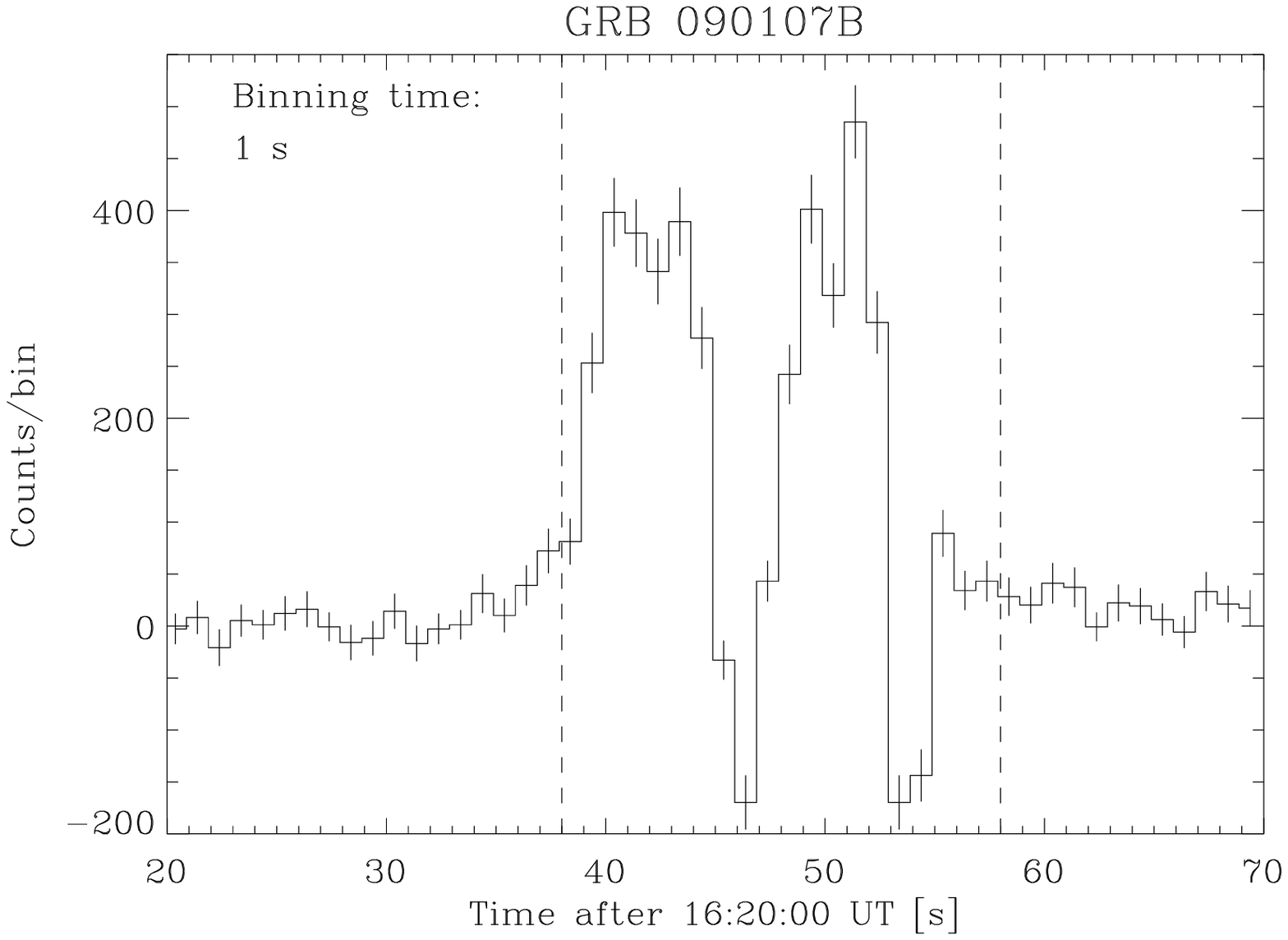}\\

  \end{tabular}
     \end{center}

      \caption{Light curves of {\it INTEGRAL} GRBs observed in the period September 2008 - February 2012. The dashed lines show the interval on which the spectral information was extracted.
              }
         \label{fig:lc1}
   \end{figure*}

  \begin{figure*}[!t]
   \begin{center}
   \begin{tabular}{cc}
   \centering

   \includegraphics[width=0.5\textwidth]{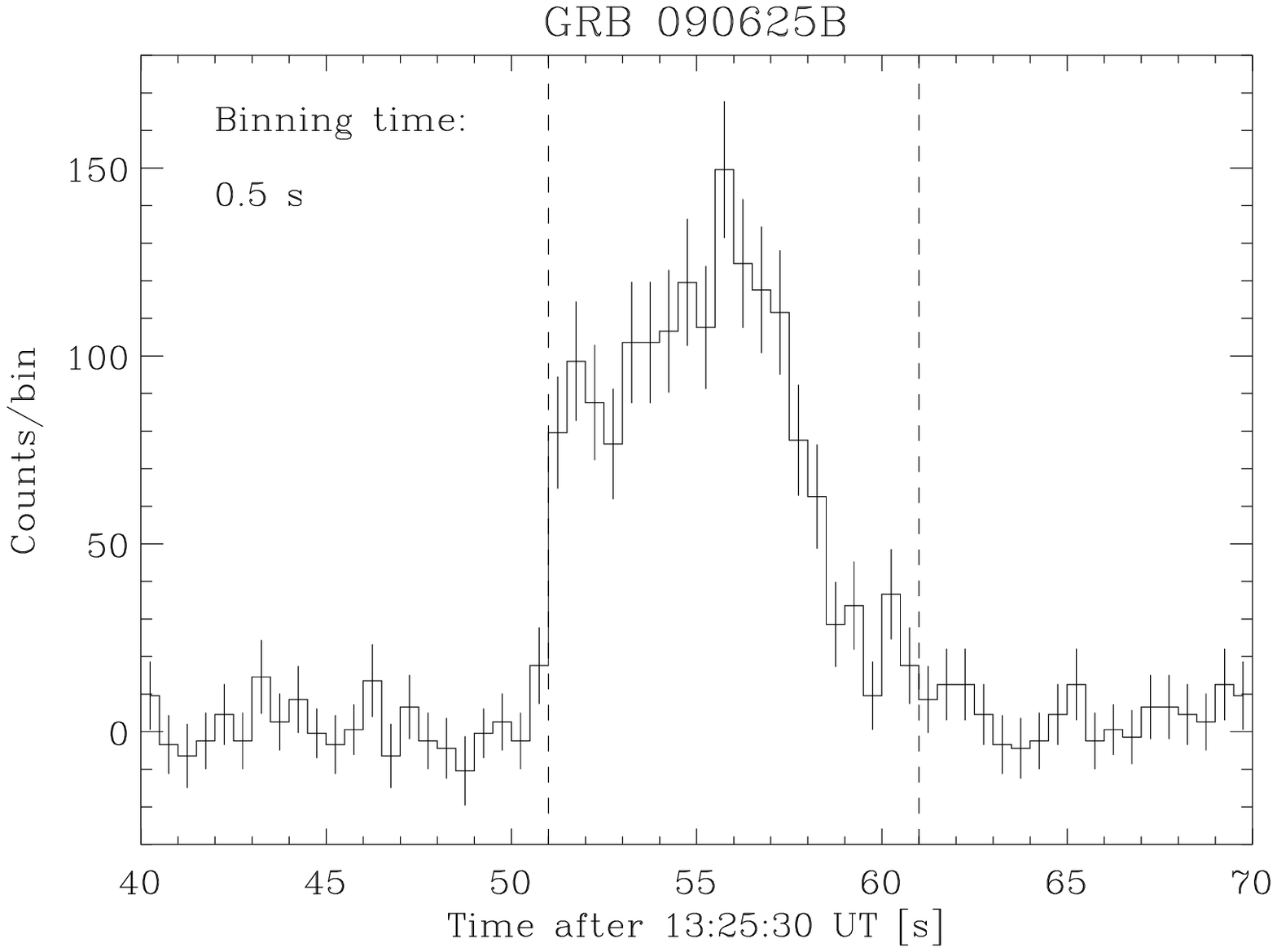}&

   \includegraphics[width=0.5\textwidth]{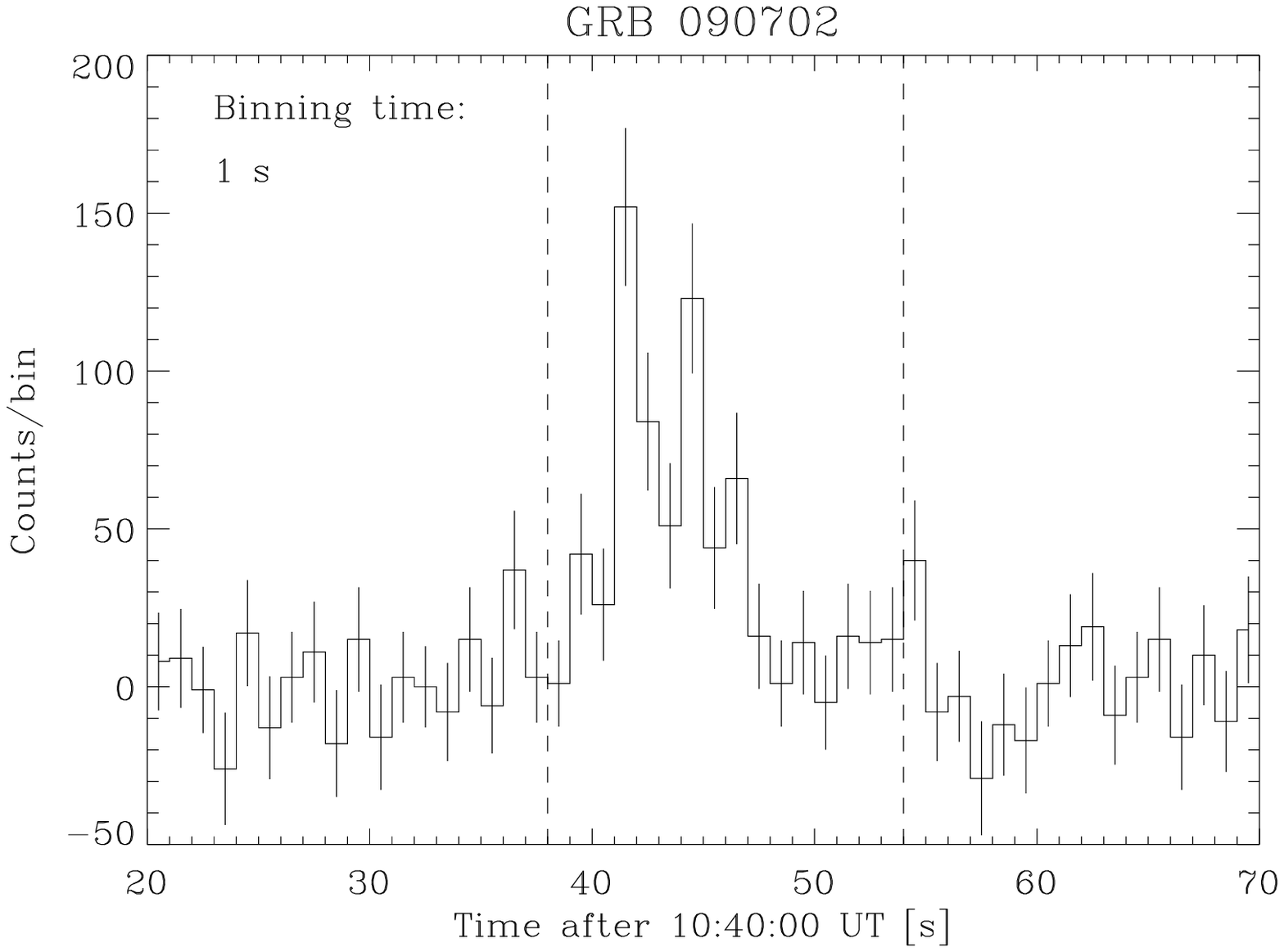}\\
   \includegraphics[width=0.5\textwidth]{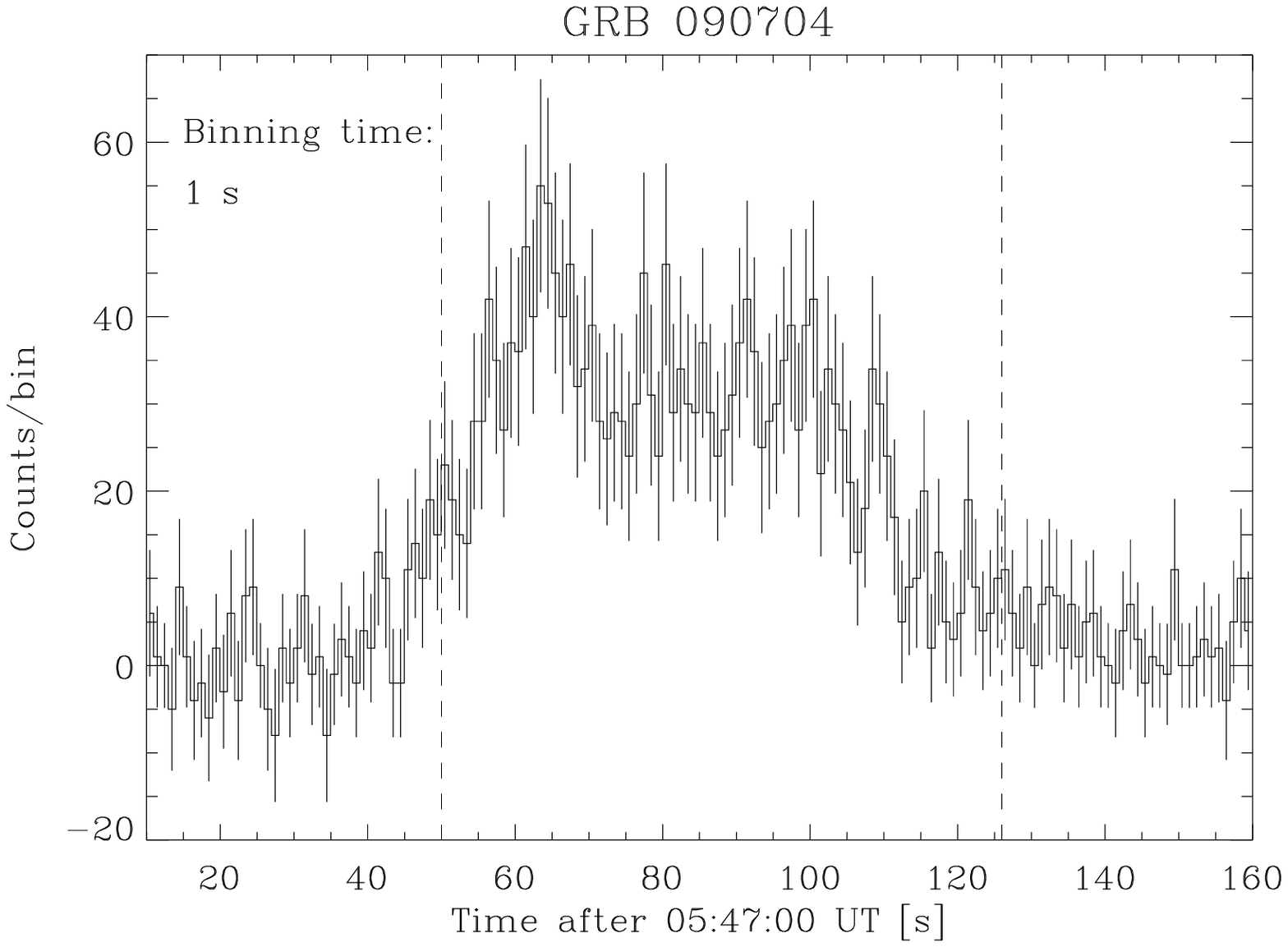}&
 
  \includegraphics[width=0.5\textwidth]{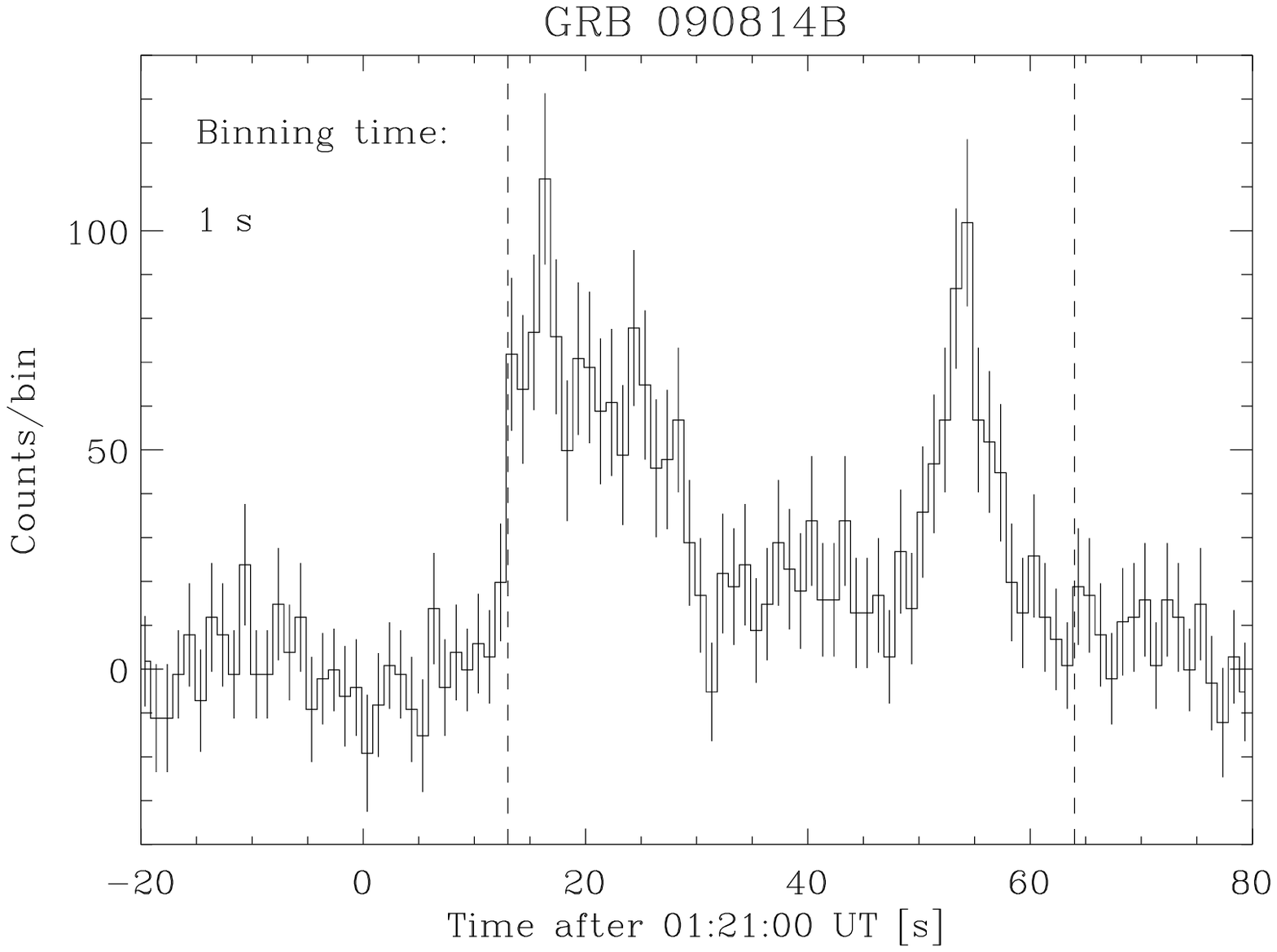}\\
   \includegraphics[width=0.5\textwidth]{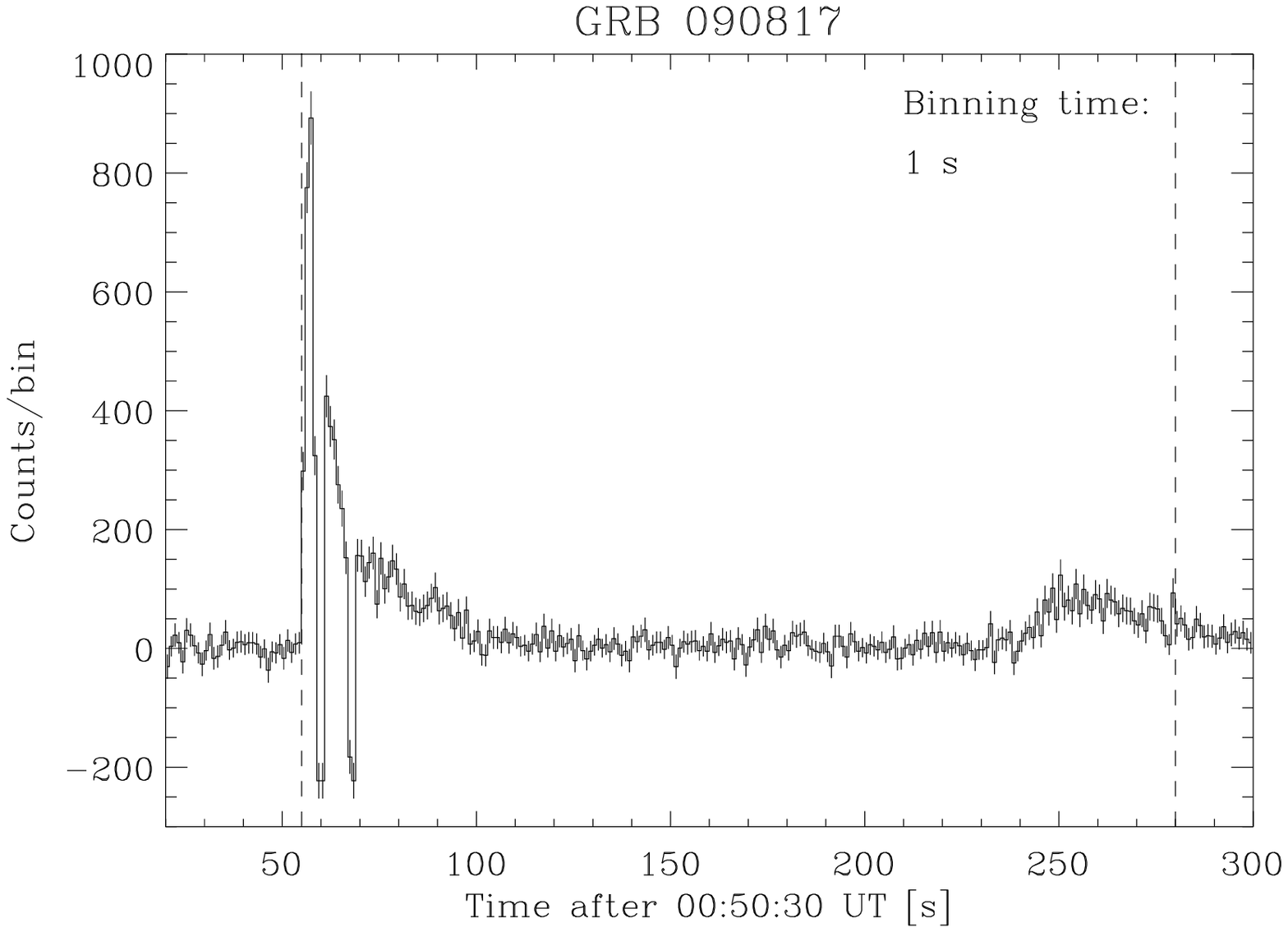}&
   \includegraphics[width=0.5\textwidth]{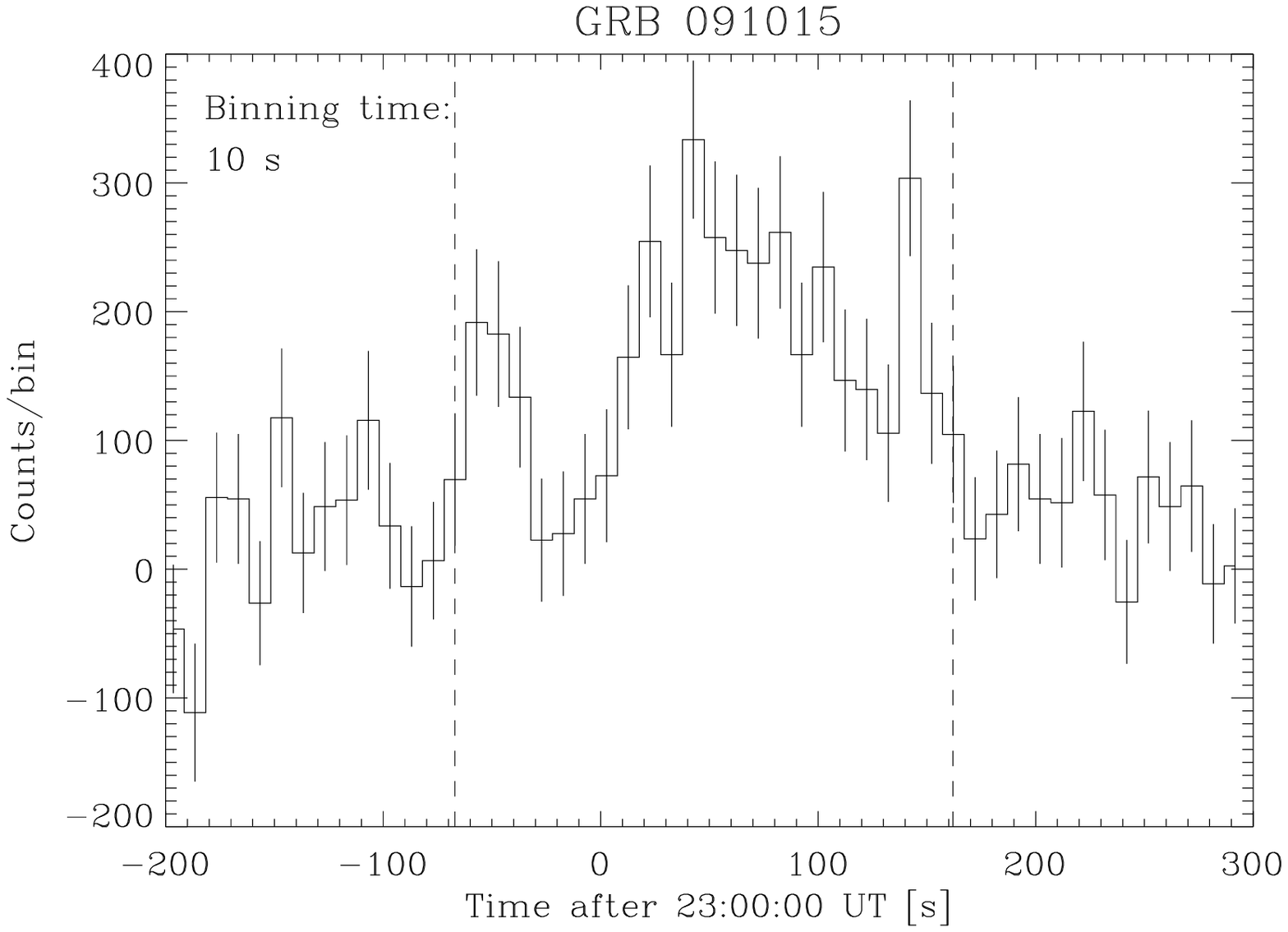}\\

  \end{tabular}
     \end{center}

      \caption{Continued.
              }
         \label{fig:lc2}
   \end{figure*}

   \begin{figure*}[!t]
   \begin{center}
   \begin{tabular}{cc}
   \centering

   \includegraphics[width=0.5\textwidth]{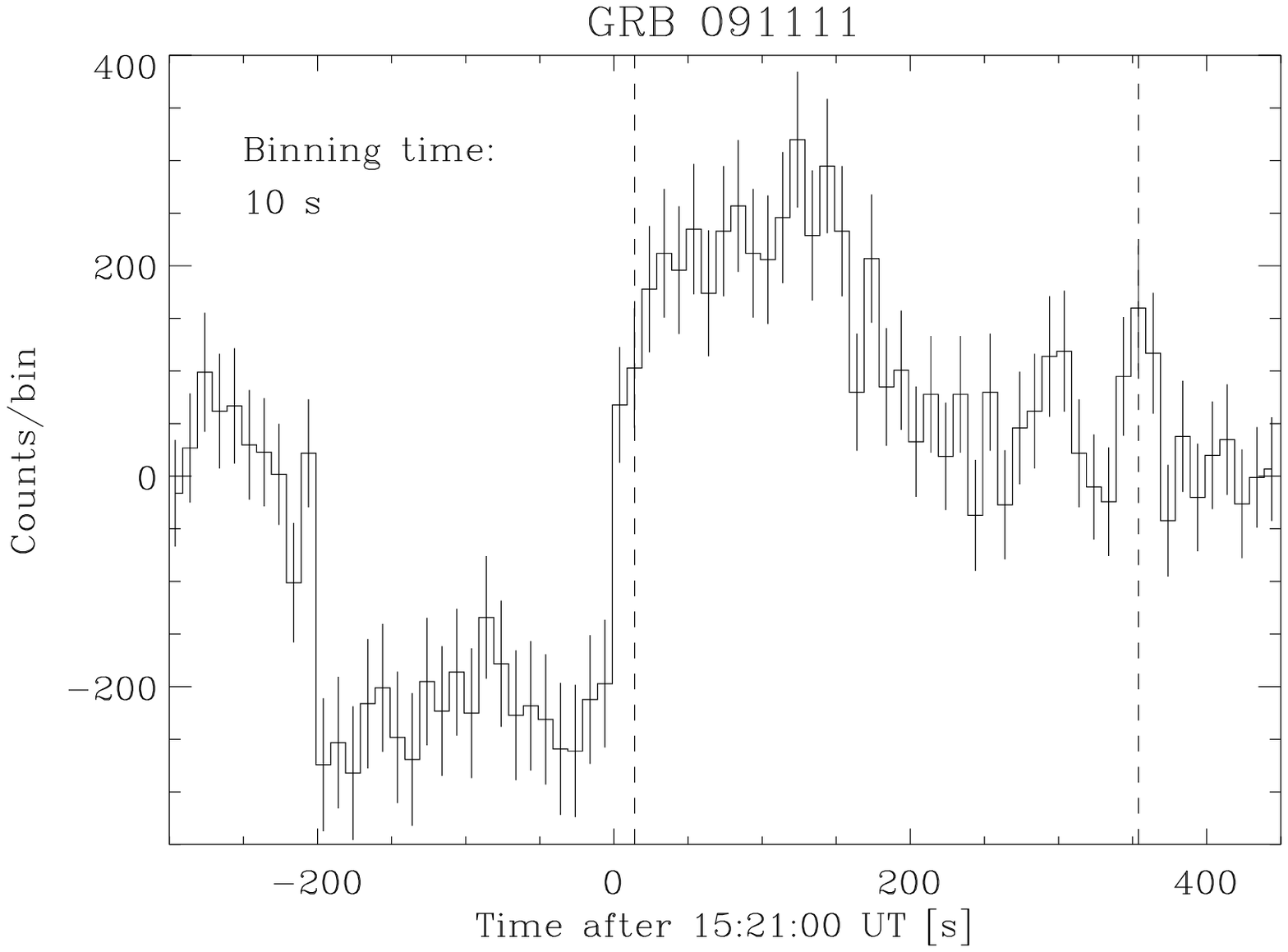}&
 
   \includegraphics[width=0.5\textwidth]{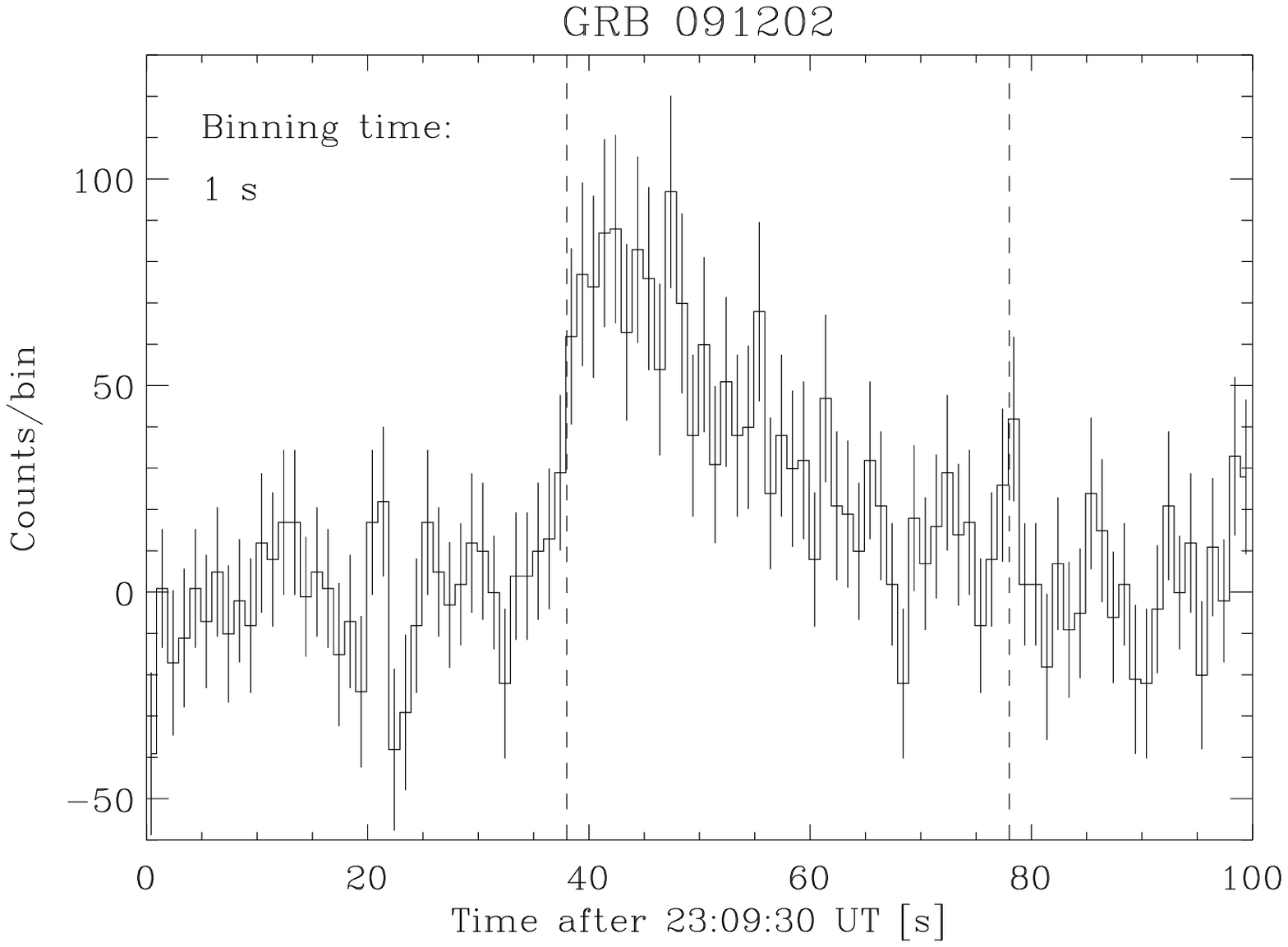}\\
   \includegraphics[width=0.5\textwidth]{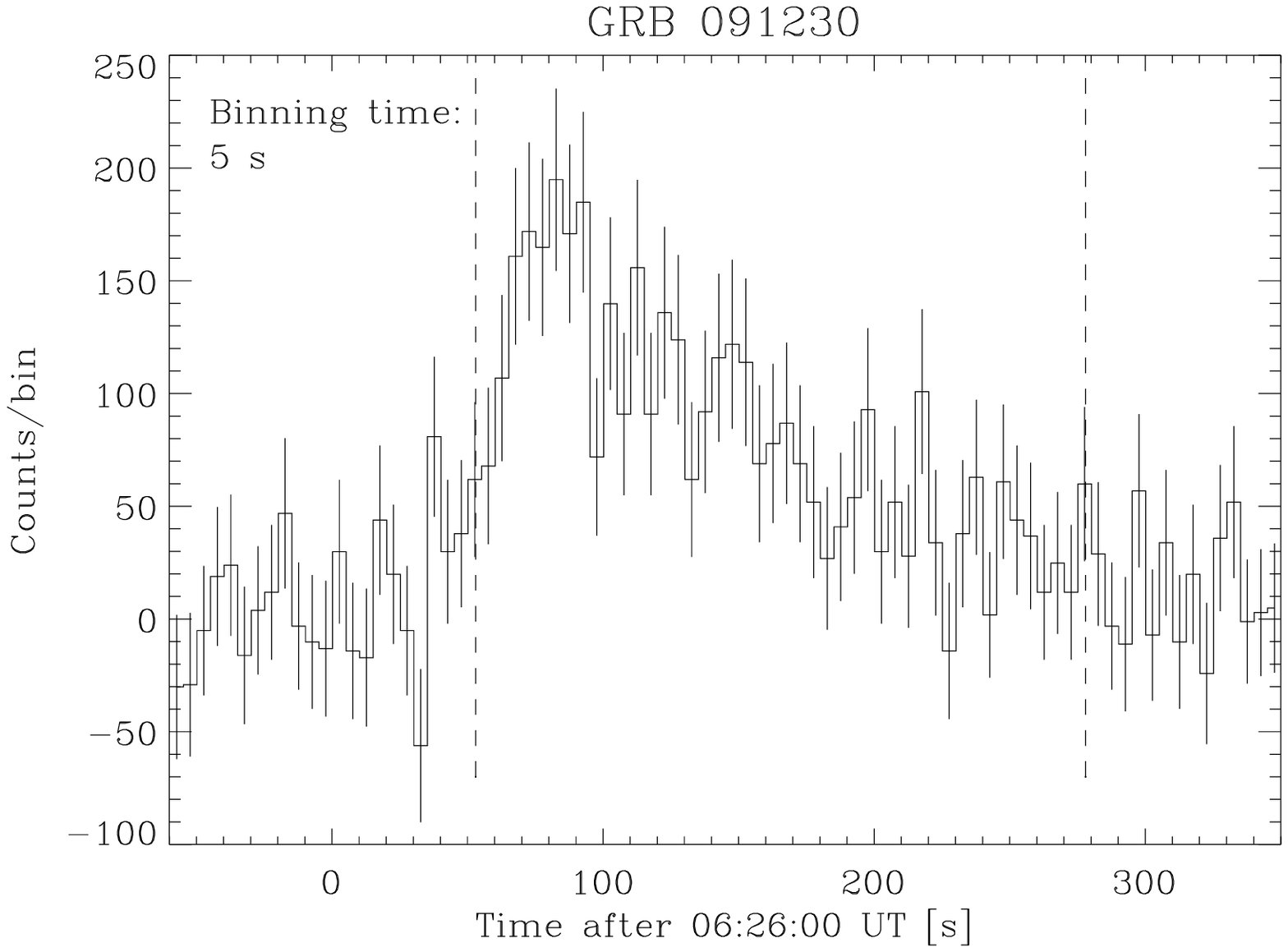}&
   \includegraphics[width=0.5\textwidth]{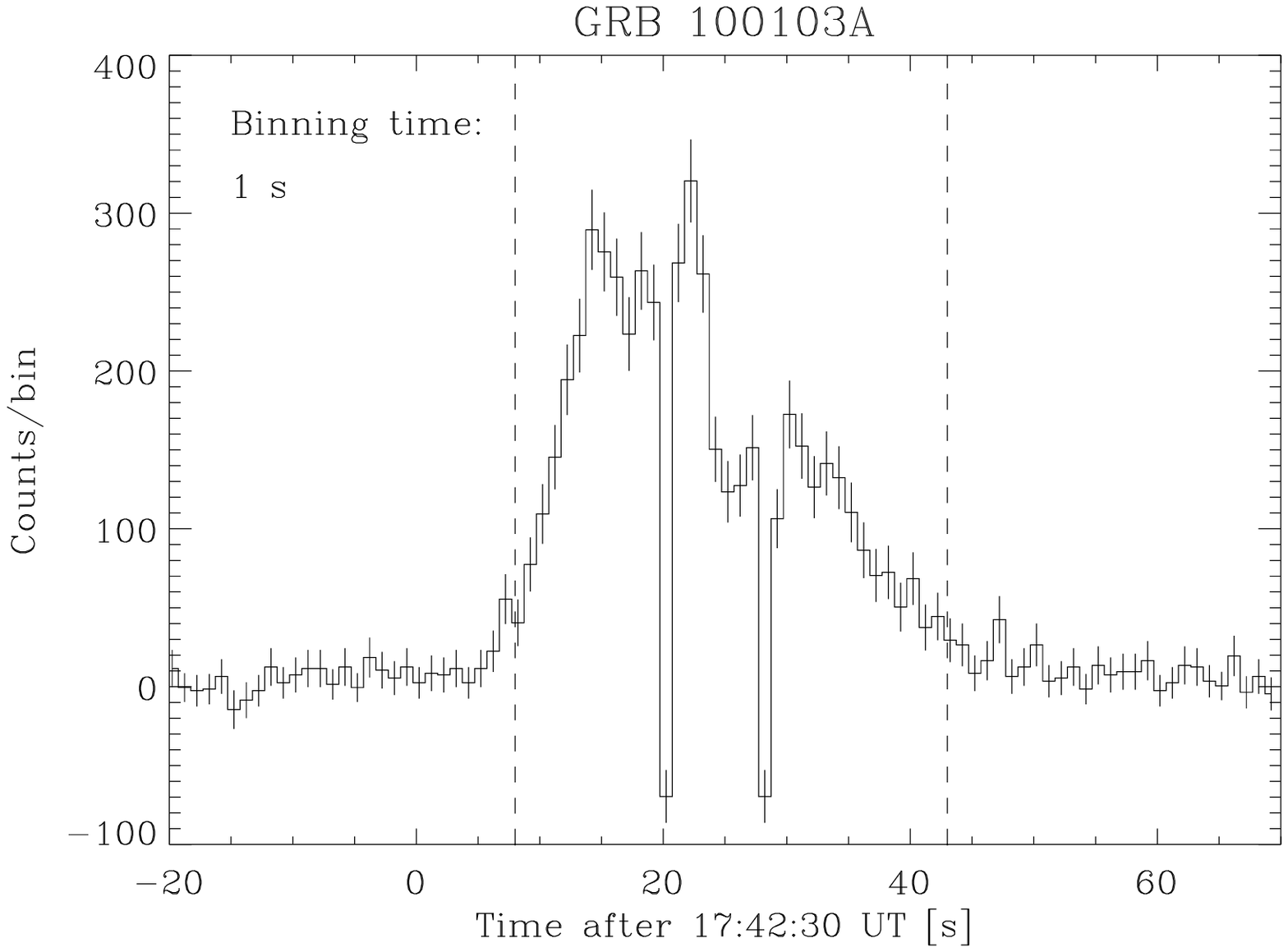}\\
   \includegraphics[width=0.5\textwidth]{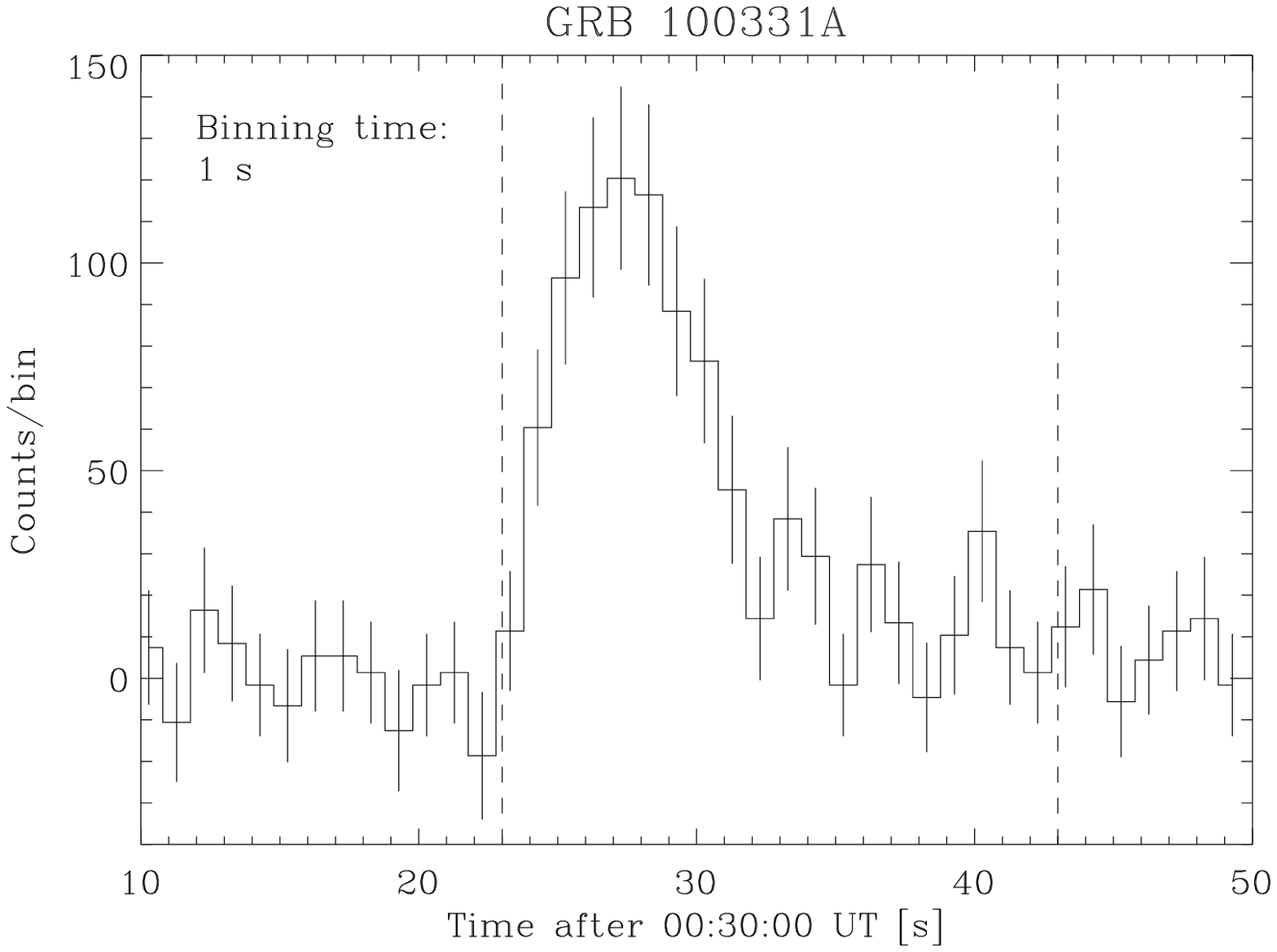}&
  \includegraphics[width=0.5\textwidth]{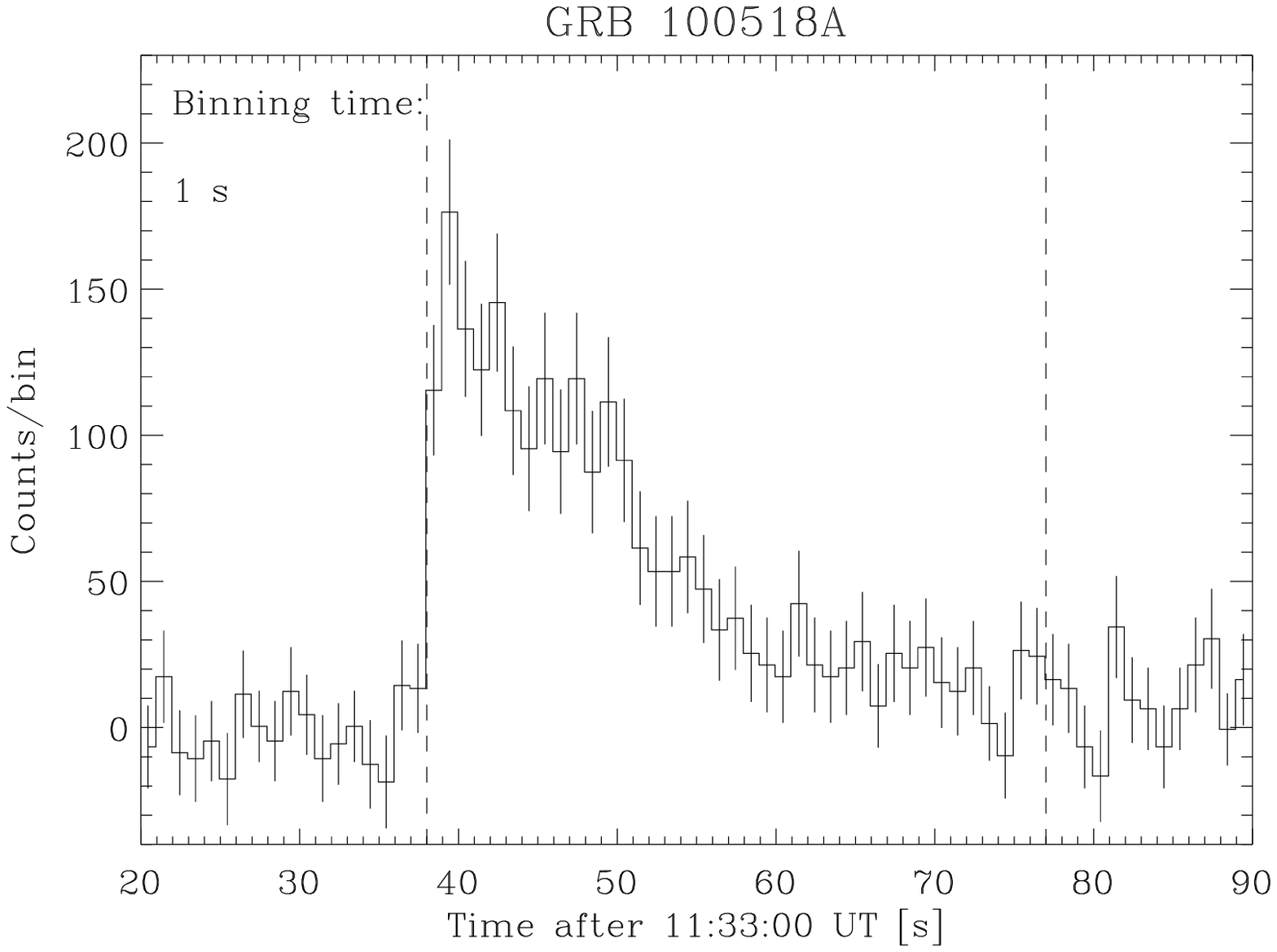}\\

  \end{tabular}
     \end{center}

      \caption{Continued.
              }
         \label{fig:lc3}
   \end{figure*}
   \begin{figure*}[!t]
   \begin{center}
   \begin{tabular}{cc}
   \centering

   \includegraphics[width=0.5\textwidth]{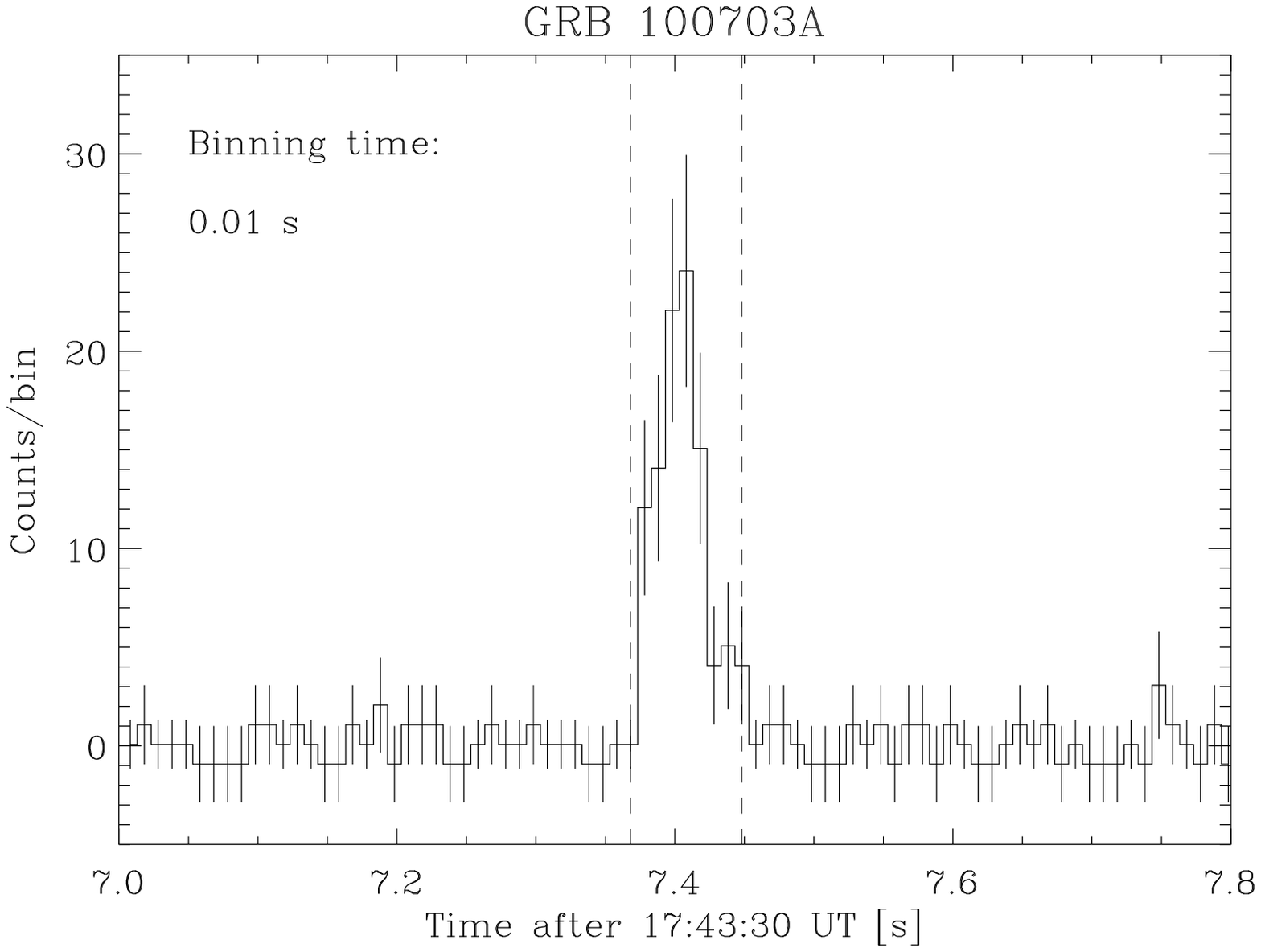}&

   \includegraphics[width=0.5\textwidth]{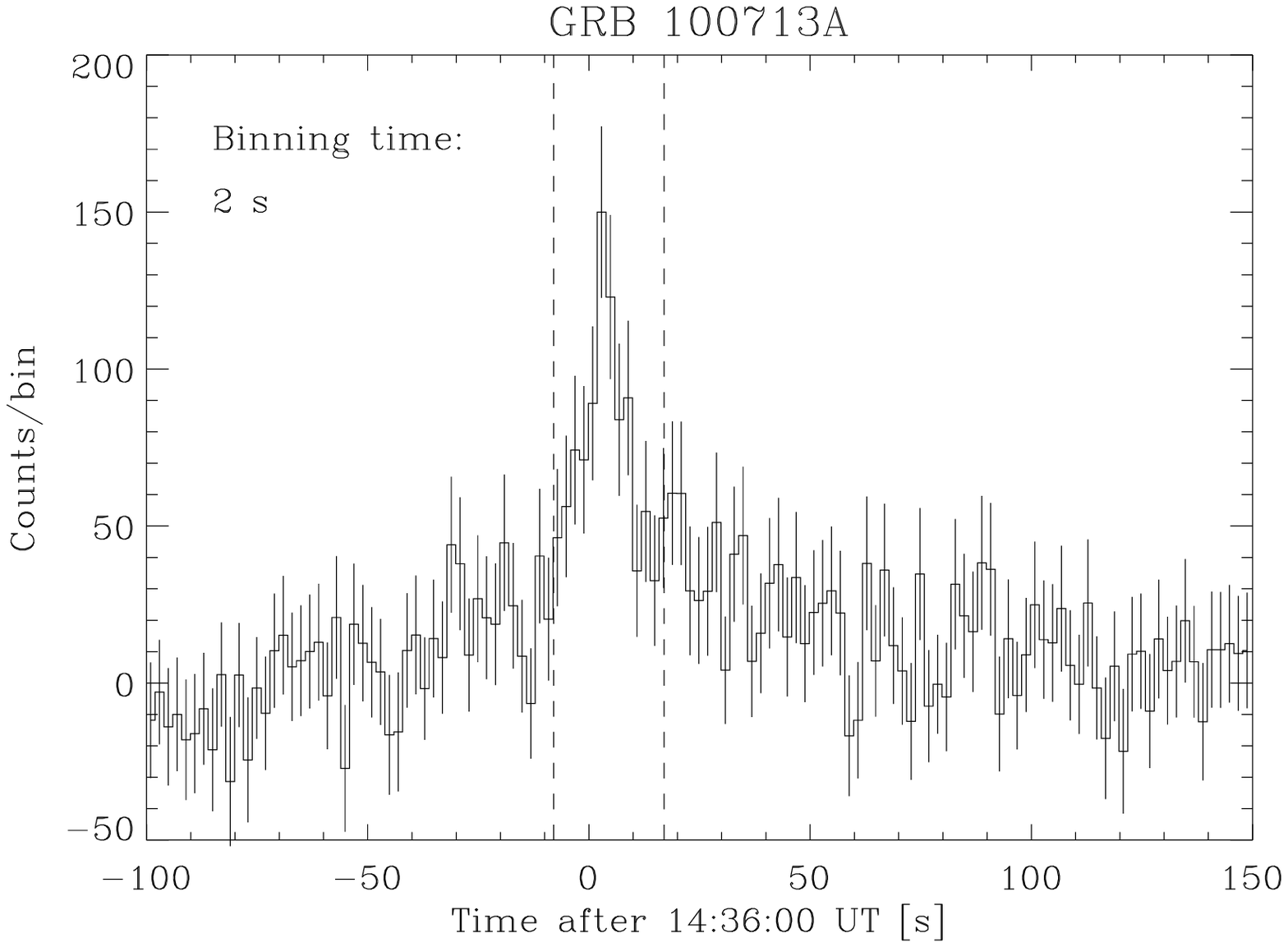}\\
   \includegraphics[width=0.5\textwidth]{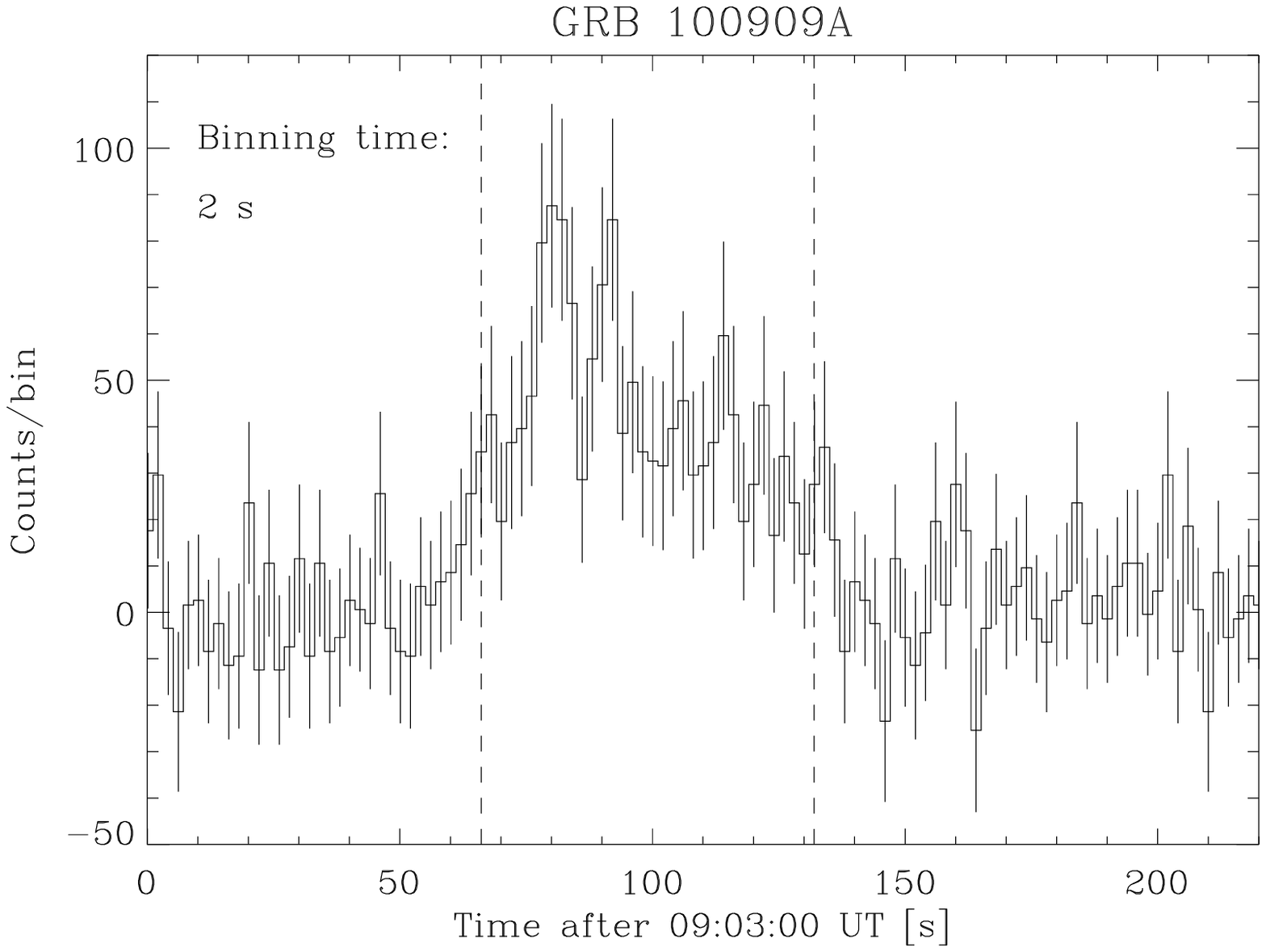}&
   \includegraphics[width=0.5\textwidth]{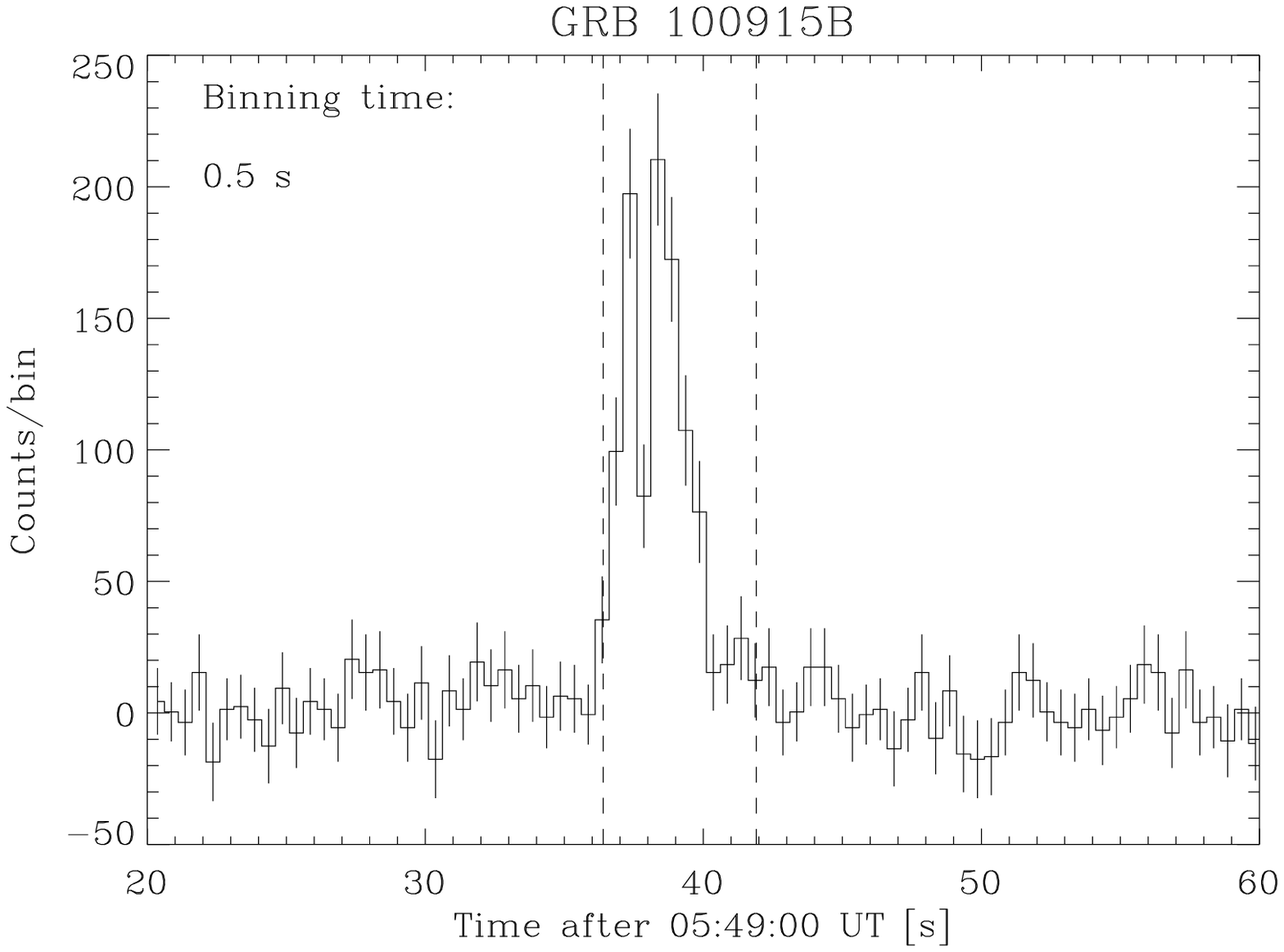}\\
   \includegraphics[width=0.5\textwidth]{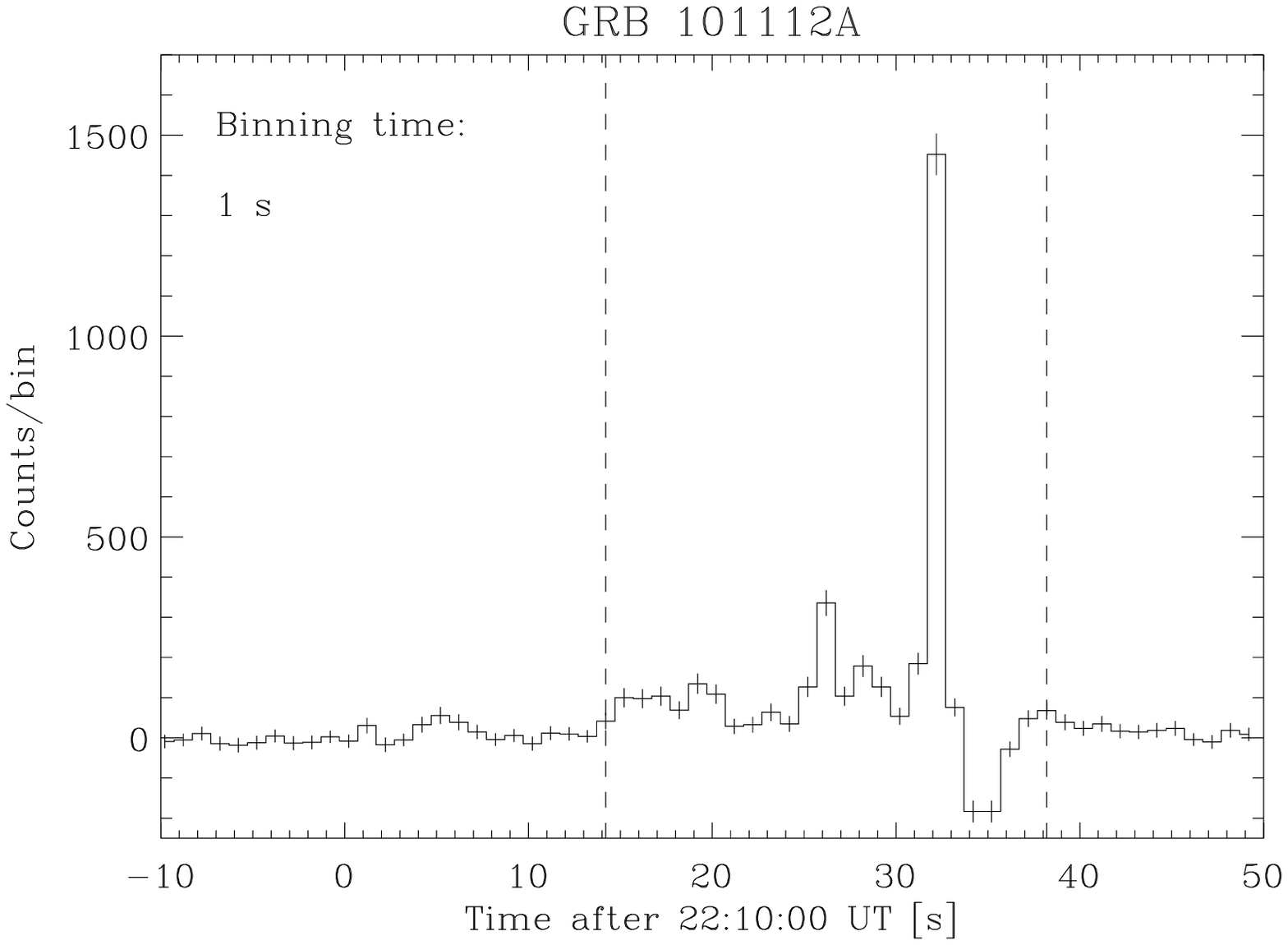}&
   \includegraphics[width=0.5\textwidth]{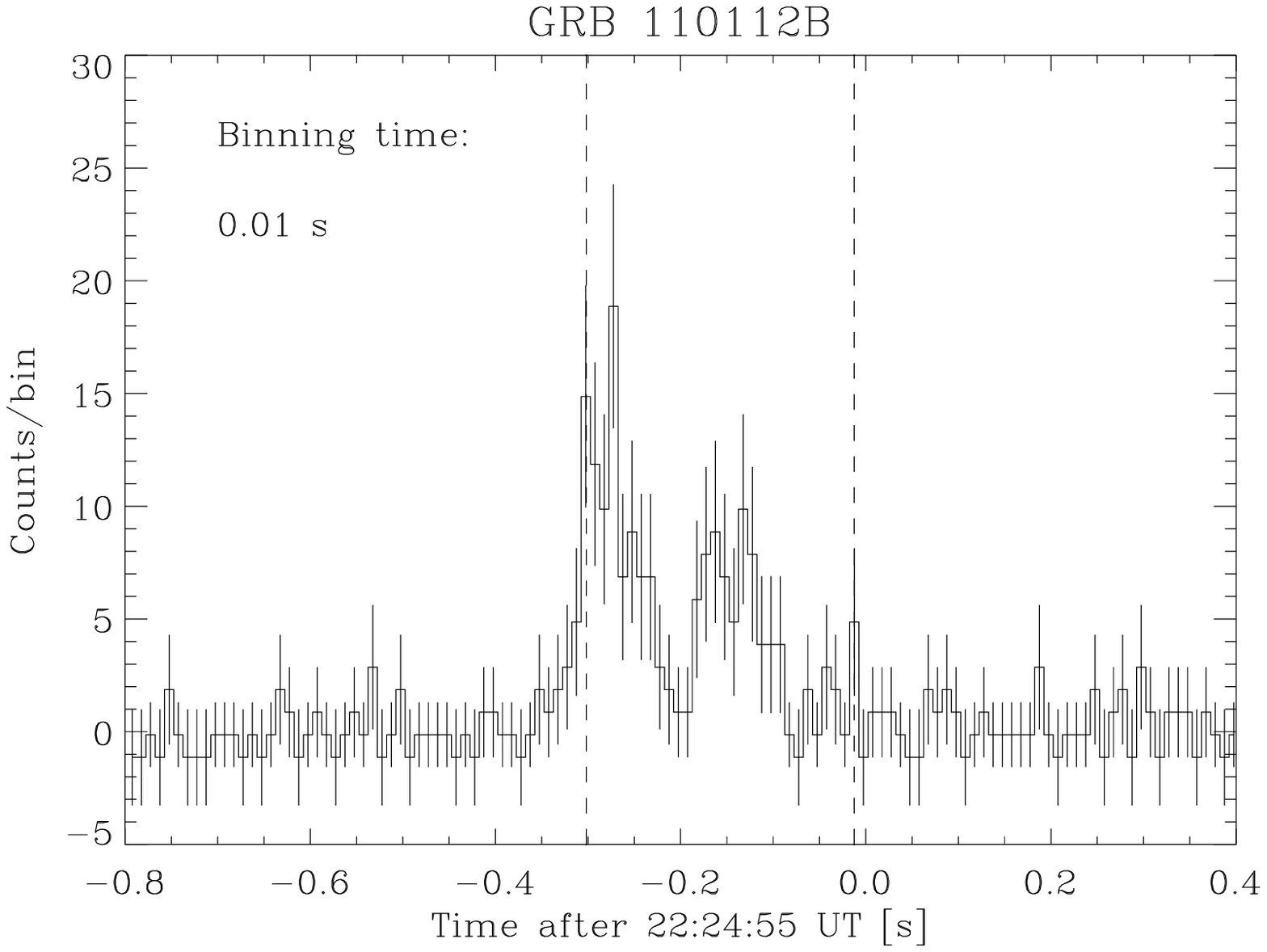}\\
  \end{tabular}
     \end{center}

      \caption{Continued.
              }
         \label{fig:lc4}
   \end{figure*}
   \begin{figure*}
   \begin{center}
   \begin{tabular}{cc}
   \centering
   \includegraphics[width=0.5\textwidth]{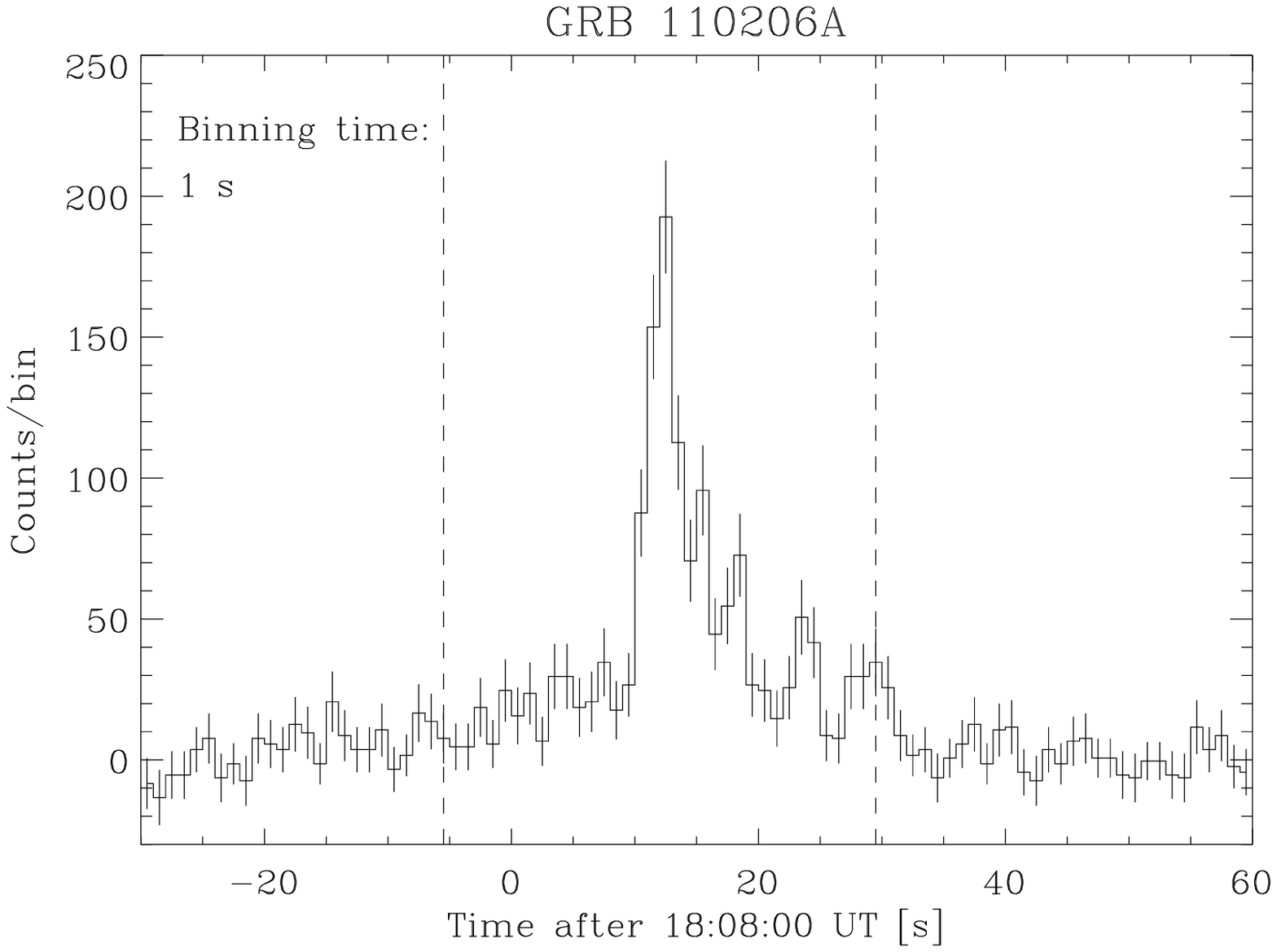}&
  \includegraphics[width=0.5\textwidth]{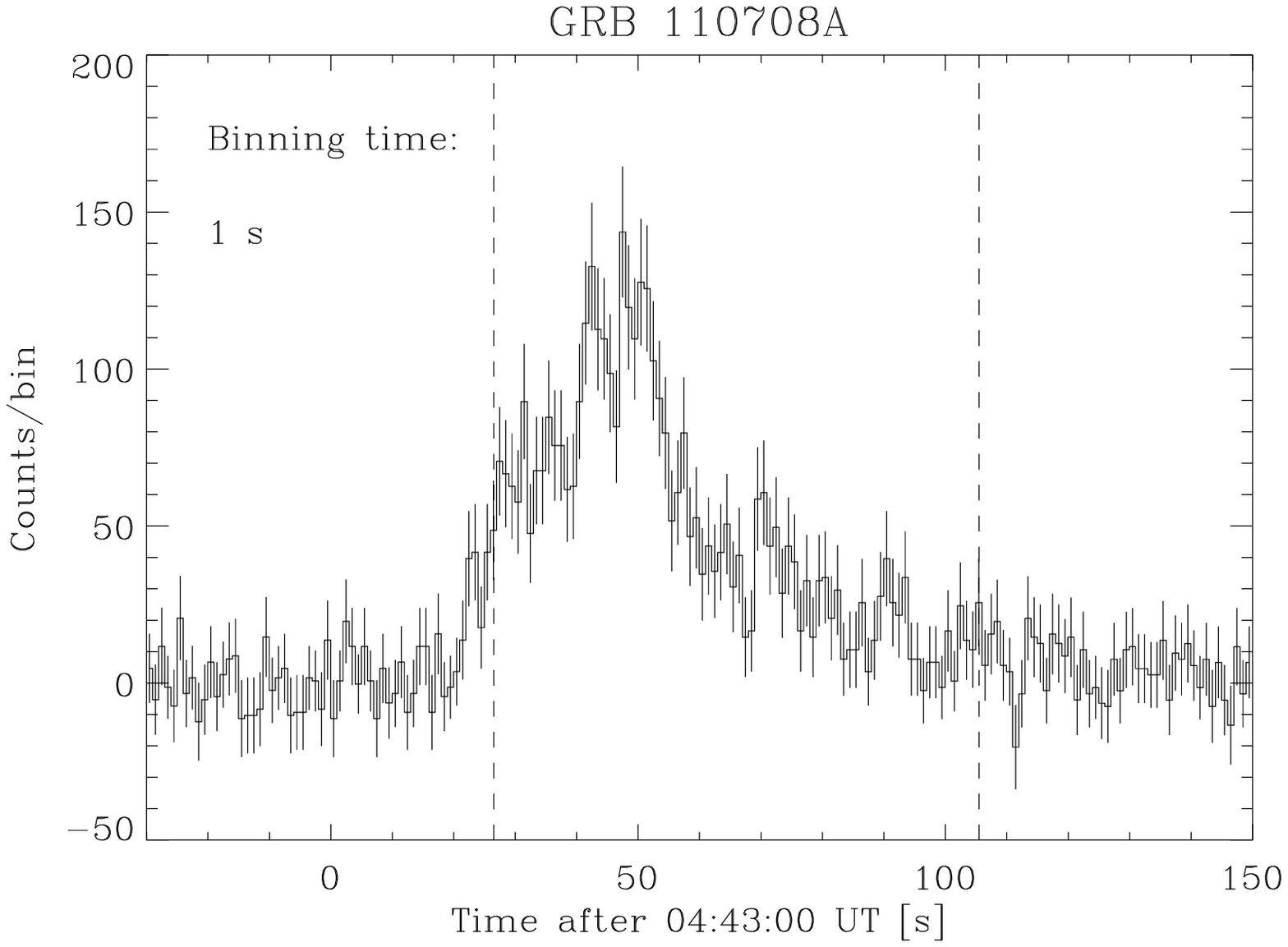}\\
  \includegraphics[width=0.5\textwidth]{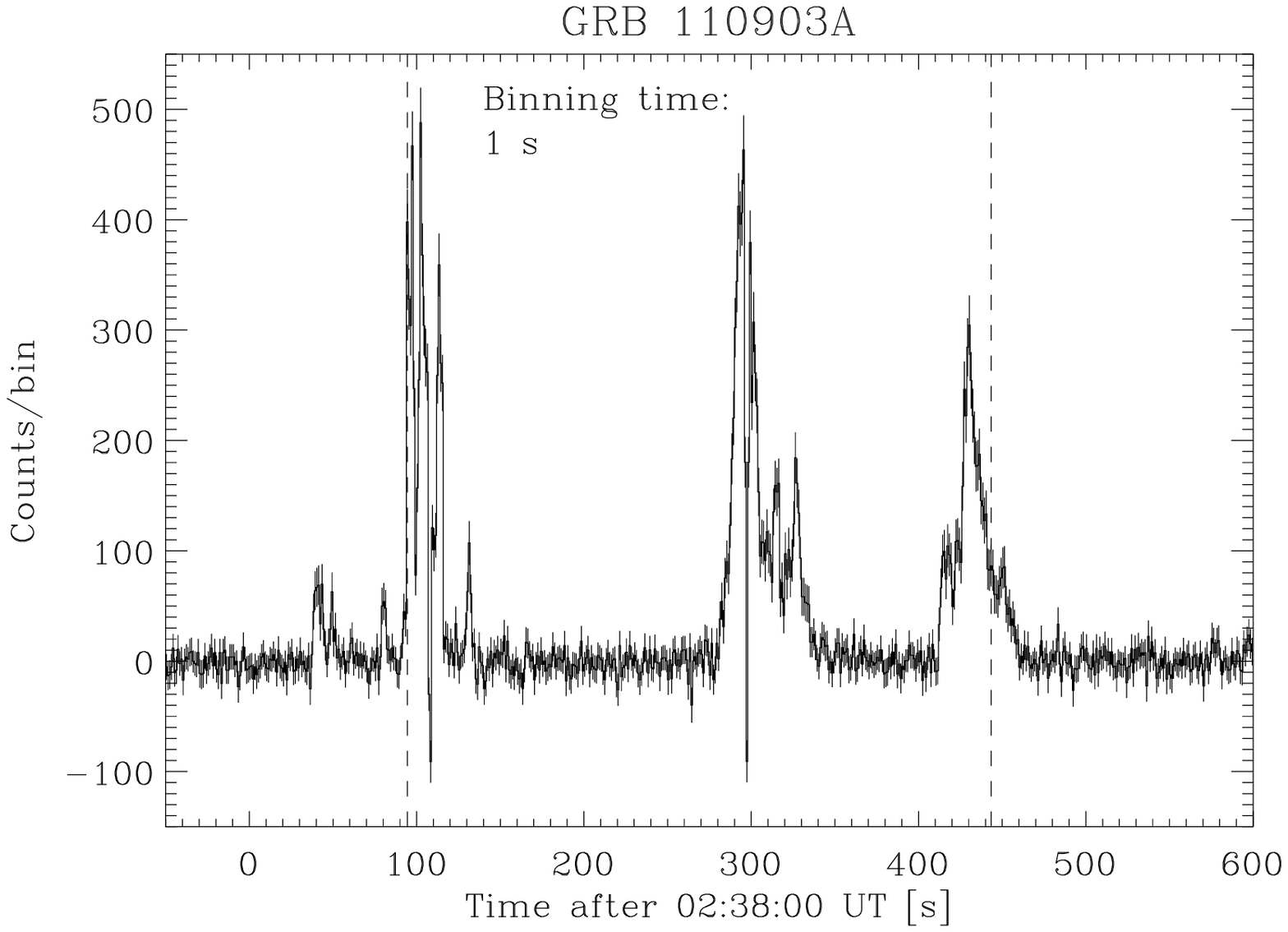}&
  \includegraphics[width=0.5\textwidth]{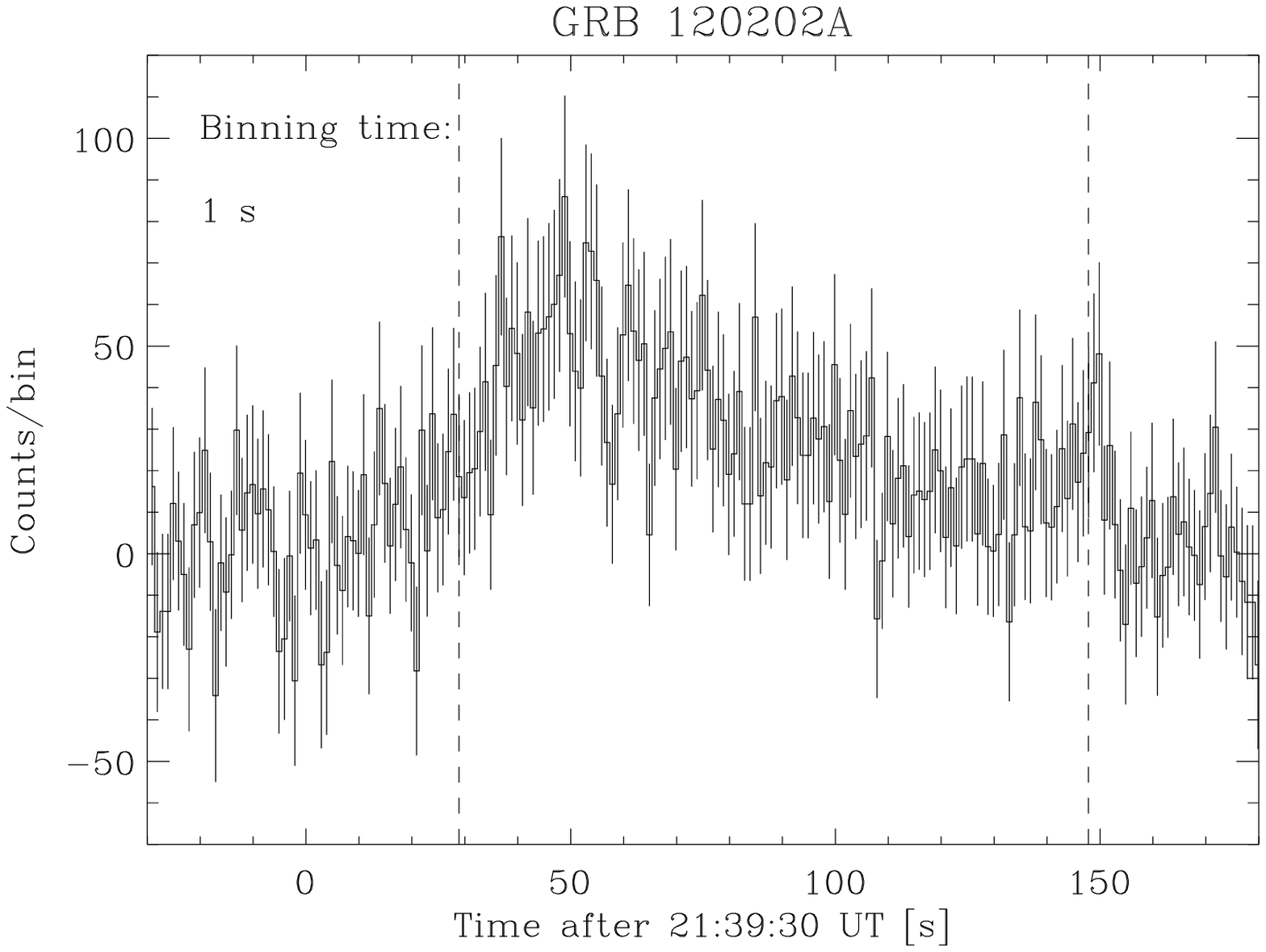}\\

  \end{tabular}
     \end{center}

      \caption{Continued.
              }
         \label{fig:lc5}
   \end{figure*}

\end{appendix}

\end{document}